\documentclass[A4,preprint,pre,10pt,twocolumn,footinbib,superscriptaddress,showpacs,showkeywords]{revtex4}
\usepackage{amsfonts}
\usepackage{amsmath}
\usepackage{amssymb}
\usepackage{graphicx}
\usepackage{float}
\usepackage{epsfig}

\begin{document}

\title[Extended canonical Monte Carlo methods]{Extended canonical Monte Carlo methods: \\Improving accuracy of microcanonical calculations using a re-weighting technique}
\author{L. Velazquez}
\affiliation{Departamento de F\'\i sica, Universidad Cat\'olica del Norte, Av. Angamos
0610, Antofagasta, Chile.}
\author{J.C. Castro-Palacio}
\affiliation{Department of Chemistry, University of Basel, Klingelbergstr. 80, 4056
Basel, Switzerland}
\keywords{Fluctuation theorems, Monte Carlo methods, Slow sampling problems}
\pacs{02.70.Tt, 05.20.Gg}
\date{\today}

\begin{abstract}
Velazquez and Curilef have proposed a methodology to extend Monte Carlo algorithms that are based on canonical ensemble. According to our previous study, their proposal allows us to overcome slow sampling problems in systems that undergo any type of temperature-driven phase transition. After a comprehensive review about ideas and connections of this framework, we discuss the application a re-weighting technique to improve the accuracy of microcanonical calculations, specifically, the well-known multi-histograms method of Ferrenberg and Swendsen. As example of application, we reconsider the study of four-state Potts model on the square lattice $L\times L$ with periodic boundary conditions. This analysis allows us to detect the existence of a very small latent heat per site $q_{L}$ during the occurrence of temperature-driven phase transition of this model, whose size dependence seems to follow a power-law $q_{L}(L)\propto(1/L)^{z}$ with exponent $z\simeq0$.$26\pm0$.$02$. It is discussed the compatibility of these results with the continuous character of temperature-driven phase transition when $L\rightarrow+\infty$.
\end{abstract}

\maketitle

\section{Introduction}

Recently \cite{vel-emc1,vel-emc2}, Velazquez and Curilef proposed a methodology that overcomes slow sampling problems due to the presence temperature driven discontinuous phase transitions (PT). Essentially, their proposal allows to improve any Monte Carlo (MC) algorithms based on canonical ensemble by introducing some suitable modifications. These extended canonical MC algorithms reduce exponential dependence of decorrelation time $\tau(N)\propto \exp(\gamma N)$ on the system size $N$ by a very weak power-law behavior $\tau(N)\propto N^{w}$. According to early estimations considering 2D $q$-state Potts models \cite{SW,pottsm,Wolff}, critical exponents $w$ of these algorithms are lower than the ones achieved using multicanonical method and its variants \cite{BergM,WangLandau,WangJStat}. Recently, we have shown that the extended canonical MC algorithms also exhibit a great performance near critical point of a temperature driven continuous PT \cite{vel-emc3}. Surprisingly, we have verified that extended version of Metropolis importance sample \cite{metro,Hastings} exhibits an efficiency slightly greater than canonical cluster algorithms of Swendsen-Wang and Wolff \cite{SW,pottsm,Wolff}.

The main goal of this work is to combine extended canonical MC algorithms with a re-weighting technique to improve the accuracy of microcanonical calculations. Information collected from different MC simulations can be combined to estimate properties at new different conditions \cite{mc3}. Specifically, we will consider \emph{multi-histograms method} of Ferrenberg and Swendsen \cite{Ferrenberg}. We shall reconsider the study of four-state Potts model on the square lattice $L\times L$ with periodic boundary conditions to improve microcanonical calculations performed in our previous work \cite{vel-emc3}. This new analysis allows us to detect the existence of a very small, but definitely non-vanishing latent heat $q_{L}$ and states with negative heat capacities $C<0$ for lattice size range of $L=22-90$, which are typical behaviors of a \emph{finite system} that undergoes a temperature driven discontinuous PT \cite{Thirring,pad,Lyn3,gro1,Moretto,Schmidt}. All associated thermodynamical behaviors, such as the entropy defect $\Delta s$ due to the region of convexity, are very small (see Fig.\ref{size.eps} below). Even using the present improvements, they are only revealed with a careful analysis of microcanonical dependencies.

At first glance, these results seem to be in contradiction with Baxter exact results \cite{Baxter}, which emphasize the continuous character of PT of this model \emph{in the thermodynamic limit} $L\rightarrow+\infty$. Anticipating our discussions on this question, we think that there is no contradiction here. Baxter exact result does not forbid the existence of negative heat capacities outside thermodynamic limit. In fact, Potts model on the square lattice $L\times L$ with $q=4$ is \emph{a marginal case} for this family of models \cite{Baxter,Creswick,Wu}, and therefore there is nothing strange if ambiguities in some thermodynamical behaviors are detected for finite lattice sizes $L$. Besides, the size dependence of our MC estimates of latent heat per site $q_{L}$ seems to follow a power-law $q_{L}(L)\propto(1/L)^{z}$ with exponent $z\simeq0$.$26$, which is fully compatible with an eventual vanishing of this quantity when $L\rightarrow+\infty$.

The paper is organized into sections as follows. Second section is devoted to discuss some important antecedents of this study. For the sake of self-consistence of the paper, we start reviewing some generalized fluctuation relations derived by Velazquez and Curilef and their relevance in MC simulations \cite{vel-tur,vel-efr,vel-geotur,vel-geft,vel-crit,vel-csm}. Afterwards, we discuss main ideas associated with extension of canonical MC methods \cite{vel-emc1,vel-emc2,vel-emc3} as well as connections with other MC methods that perform microcanonical calculations \cite{mc3}. Third section is devoted to discuss application of multi-histograms method to improve this type microcanonical MC calculations. As example of application, we discuss the improvement of microcanonical estimations of four-state Potts model on the square lattice $L\times L$ with periodic boundary conditions. Final remarks and open questions are discussed in the fourth section.

\section{Antecedents}

\subsection{Generalized fluctuation relations and their application to MC simulations}

\begin{figure}[tbp]
\begin{center}
\includegraphics[width=3.5in]{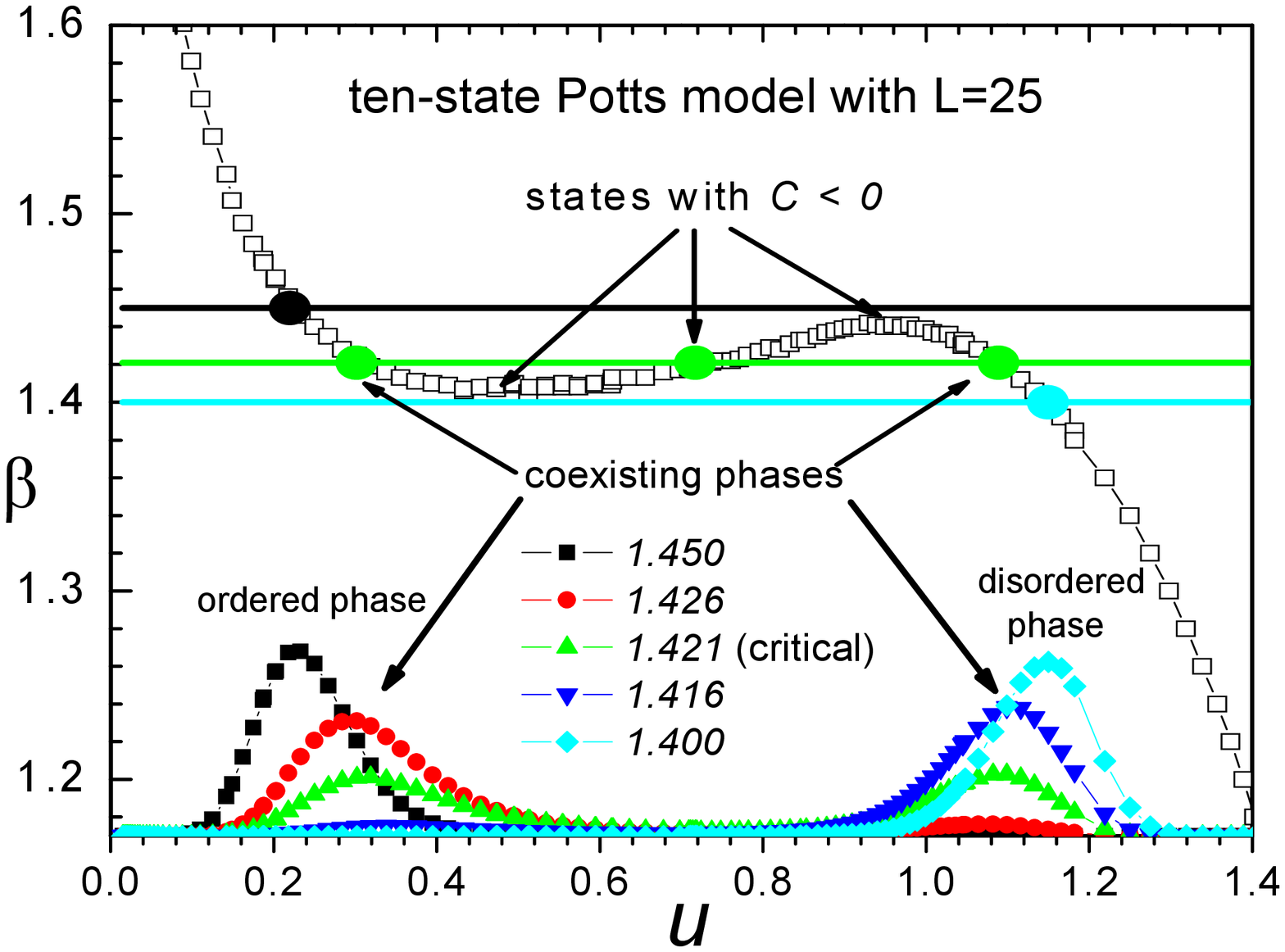}
\end{center}
\caption{(Color online) Behavior of energy distributions within canonical ensemble (\ref{can}) along the occurrence of phase coexistence phenomenon of the ten-state Potts model on the square lattice $25 \times 25$ with periodic boundary conditions ($u = U/N$ is the energy per site) (after \cite{vel-tur}). This study clearly illustrates bimodal character of energy distribution functions when the inverse temperature parameter $\beta$ of canonical ensemble takes values around the critical value $\beta_{c}\simeq 1$.$421$ of temperature driven discontinuous PT. Notice that the branch of \emph{microcanonical caloric curve} $\beta(u)=\partial s(u)/\partial u$ (open squares) with states with negative heat capacities $C<0$ is poorly populated by using a bath with constant temperature, since these states are canonically unstable. The values of bath inverse temperature $\beta$ are represented here by horizontal lines. Intersection points of these horizontal lines with microcanonical inverse temperature correspond to the energies where energy distribution function exhibits its local maxima and minima.}\label{potts.eps}
\end{figure}

Since early demonstration of generalized fluctuation relation:
\begin{equation}\label{fdr}
C=\beta^{2}\left\langle\delta U^{2}\right\rangle+C\left\langle\delta \beta_{\omega}\delta U\right\rangle
\end{equation}
by Velazquez and Curilef, it was clearly evidenced that its associated background conditions of derivation can be employed to extend any MC algorithm based on canonical ensemble:
\begin{equation}\label{can}
\omega_{c}(U|\beta)=\frac{1}{Z(\beta)}\exp\left(-\beta U\right)
\end{equation}
(see in subsection 3.1 in Ref.\cite{vel-tur}). As early shown by Boltzmann and Gibbs \cite{Gibbs}, canonical ensemble (\ref{can}) describes a system of interest that is put in thermal contact with an environment of constant temperature, or equivalently, a thermal bath of infinite heat capacity. In fully analogy as the known relation \cite{Reichl}:
\begin{equation}\label{can.fr}
C=\beta^{2}\left\langle\delta U^{2}\right\rangle
\end{equation}
of classical fluctuation theory is employed in any MC study based on canonical ensemble (\ref{can}) to obtain the heat capacity $C$ from the energy fluctuations, the more general fluctuation relation (\ref{fdr}) can be employed with the same purpose in any MC study where the environmental inverse temperature $\beta_{\omega}$ experiences thermal fluctuations that are coupled with thermal fluctuations of the system energy $U$ \cite{vel-tur,vel-efr,vel-geotur}.

\begin{figure}[tbp]
\begin{center}
\includegraphics[width=3.5in]{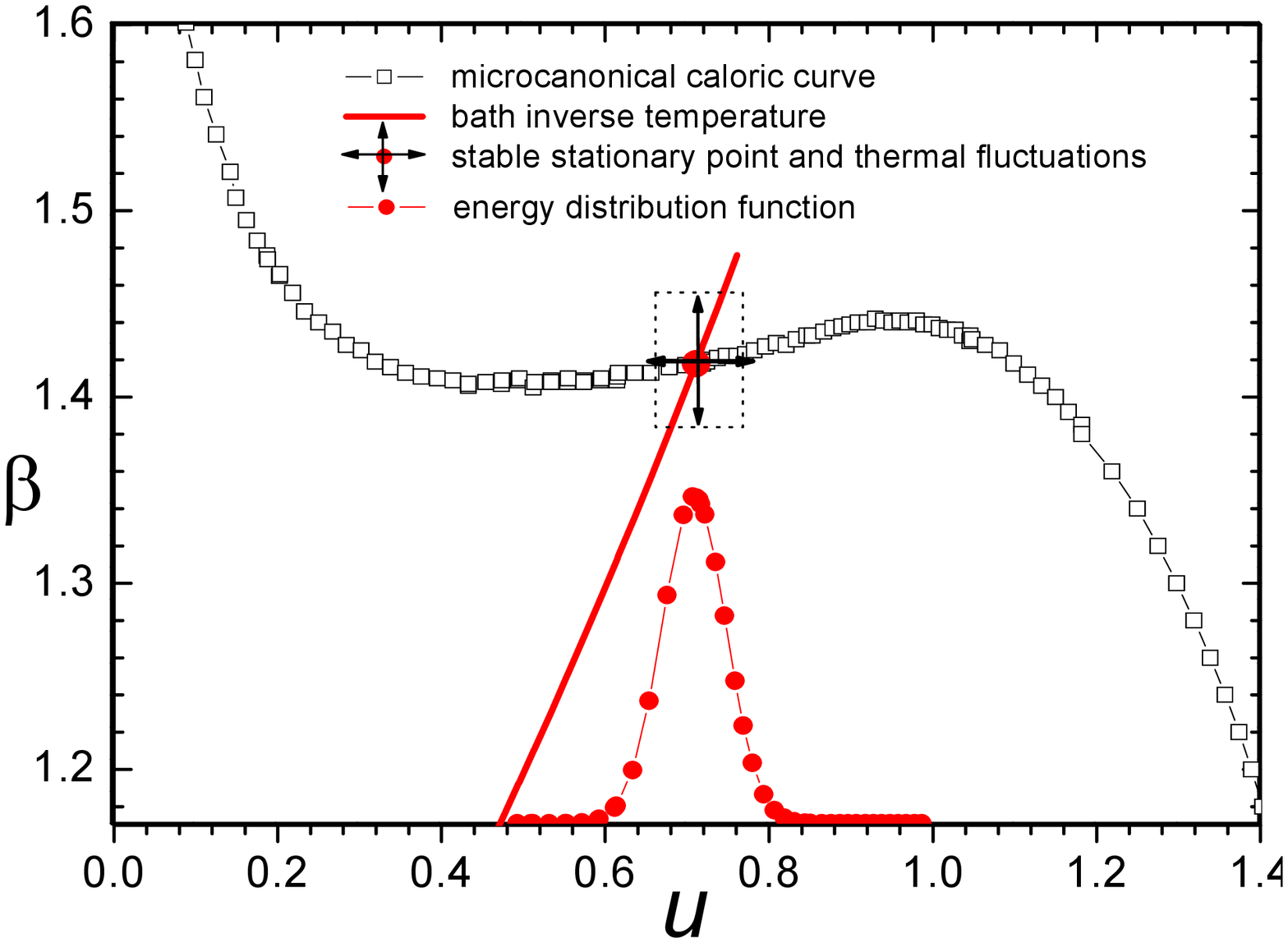}
\end{center}
\caption{(Color online) The use of a bath with a finite heat capacity $C_{\omega}$ enables a direct study of the branch of microcanonical caloric curve (open squares) with negative heat capacities, which is shown for the same model system of Fig.\ref{potts.eps} (after \cite{vel-tur}). Here, energy distribution exhibits a single Gaussian-peak that is located inside the region where microcanonical caloric curve exhibits negative heat capacities $C<0$. Both the bath inverse temperature $\beta_{\omega}$ (thick red line) and the system energy $U$ exhibit thermal fluctuations around to their equilibrium values (coordinates of the red circle that marks the interception point between microcanonical caloric curve and the bath inverse temperature curve). In the past \cite{Gerling}, Gerling and H\"{u}ller proposed this type of arguments to obtain microcanonical caloric curve considering the expectation values $\left\langle\beta_{\omega}\right\rangle$ and $\left\langle U\right\rangle$. The analysis their thermal fluctuations $\left\langle\delta U^{2}\right\rangle$ and $\left\langle\delta \beta_{\omega}\delta U\right\rangle$ enables a direct derivation of a negative value of microcanonical heat capacity $C$ at the equilibrium energy considering fluctuation relation (\ref{fdr}).}\label{gencanonical.eps}
\end{figure}

A simple realization of this effect arises when the system of interest is put in thermal contact with a bath with finite heat capacity $C_{\omega}$. The inverse temperature $\beta_{\omega}$ of the bath will not longer a constant parameter as the case of canonical ensemble (\ref{can}). On the contrary, it turns a \emph{dynamical variable} that evolves as a consequence of the underlying thermodynamic interaction, which is described in Eq.(\ref{fdr}) by the existence of a non-vanishing correlation function $\left\langle\delta \beta_{\omega}\delta U\right\rangle$. For any MC study based on the consideration of a bath with finite heat capacity, its corresponding inverse temperature $\beta_{\omega}$ is a dynamical variable that evolves during the course of simulation. It is noteworthy that these same arguments were employed in the past by Gerling and H\"{u}ller to proposed the so-called \emph{dynamic ensemble method} \cite{Gerling}. Velazquez and Curilef methodology to extend canonical MC algorithms could be regarded as an improvement of Gerling and H\"{u}ller proposal \cite{vel-emc3}. This methodology now includes modifications that enhance potentialities of this type of formalism, which also share several connections with some ideas proposed in the past by Challa and Hetherington \cite{Challa,Hetherington,ChallaPRA}.

An advantage of this perspective is that it involves \emph{a stronger control on the system fluctuating behavior and its stability} than the one considered by canonical ensemble (\ref{can}). Eq.(\ref{fdr}) is compatible with the existence of negative heat capacities $C<0$ that appear during the occurrence of a temperature driven discontinuous PT \cite{Thirring,pad,Lyn3,gro1,Moretto,Schmidt}. This fact is easy to see by rephrasing Eq.(\ref{fdr}) as follow:
\begin{equation}
C\left[1-\left\langle\delta \beta_{\omega}\delta U\right\rangle\right]=\beta^{2}\left\langle\delta U^{2}\right\rangle,
\end{equation}
where the pre-requisite of negative heat capacity $C<0$ implies the inequality $\left\langle\delta \beta_{\omega}\delta U\right\rangle>1$. Clearly, the study of systems with this behavior is not possible for MC simulations based on canonical ensemble (\ref{can}), where thermal fluctuations of bath inverse temperature $\delta \beta_{\omega}\equiv0$. In fact, its associated fluctuation relation (\ref{can.fr}) is compatible with positive heat capacities only. The presence of states with negative heat capacity can be manifested by the \emph{multimodal character} of energy distribution function within canonical ensemble \cite{vel-emc1}. This mathematical behavior of canonical energy distributions is shown in Fig.\ref{potts.eps} for the case of ten-state Potts model on the square lattice. It is noteworthy that states with negative heat capacity associated with S-bend of microcanonical caloric curve $\beta(u)$ are poorly populated within canonical ensemble. Such anomalous states can be studied in a MC simulation that implements the existence of a thermal contact with bath of finite heat capacity, which is shown in Fig.\ref{gencanonical.eps} for the same model system \cite{vel-tur}.

Recently \cite{vel-emc3}, we have emphasized that the present arguments can be useful in MC studies of systems that undergo a temperature driven continuous PT. As discussed elsewhere \cite{Reichl}, heat capacity $C$ can be very large, or even diverge, when a system approaches critical point of a temperature driven continuous PT. According to canonical fluctuation relation (\ref{can.fr}), a divergence of the heat capacity $C$ implies a divergence of energy fluctuations $\left\langle\delta U^{2}\right\rangle$. In MC simulations, large fluctuations imply large configurational changes that are also accompanied of slow sampling problems \cite{mc3}. Commonly, the strategy to overcome these difficulties is the implementation of non-local MC moves, namely, the use of \emph{clusters MC algorithms} \cite{SW,pottsm,Wolff}. By itself, fluctuation relation (\ref{fdr}) suggests an alternative way to face these problems: the use of a bath with positive finite heat capacity $C_{\omega}$.

For a simple illustration of the above idea, let us consider the first-order approximation for thermal fluctuations of bath inverse temperature, $\delta \beta_{\omega}=-\beta^{2}_{\omega}\delta U_{\omega}/C_{\omega}\equiv\beta^{2}\delta U/C_{\omega}$, which enables us to rephrase fluctuation relation (\ref{fdr}) as follows:
\begin{equation}\label{derived.fr}
\frac{CC_{\omega}}{C+C_{\omega}}=\beta^{2}\left\langle\delta U^{2}\right\rangle.
\end{equation}
Accordingly, the system energy fluctuations are fully determined by the bath heat capacity $C_{\omega}$ when the system heat capacity $C\rightarrow+\infty$:
\begin{equation}
C_{\omega}=\beta^{2}\left\langle\delta U^{2}\right\rangle\equiv\beta^{2}\left\langle\delta U^{2}_{\omega}\right\rangle.
\end{equation}
It is easy to realize that this last result is fully equivalent to canonical relation (\ref{can.fr}) when one permutes the roles of the bath and the system of interest. The positivity of right side of Eq.(\ref{derived.fr}) also implies that the study of a system with negative heat capacity $C<0$ demands the fulfilment of the following inequality:
\begin{equation}\label{Thirring.ineq}
C_{\omega}<\left|C\right|,
\end{equation}
which was pioneering derived by Thirring in Ref.\cite{Thirring}. These reasonings show that heat capacity $C_{\omega}$ of the bath should not be finite only, but also it must satisfy the above constraint. Even, the value of heat capacity $C_{\omega}$ can be optimized to reduce as low as possible the statistical uncertainties associated with determination of the microcanonical caloric curve of the system of interest (see Eq.(\ref{optimal.heatbath}) below). Fluctuation relation (\ref{derived.fr}) was also derived by Challa and Hetherington in Ref.\cite{ChallaPRA} using different arguments.

Energy-temperature fluctuation relation (\ref{fdr}) is just a particular case of more general fluctuation theorems \cite{vel-geft,vel-crit}. As example, the following fluctuation relation \cite{Reichl}:
\begin{equation}\label{suscept.fr}
\chi_{T}=\beta\left\langle \delta M^{2}\right\rangle
\end{equation}
is also widely employed in MC simulations to obtain isothermal magnetic susceptibility $\chi_{T}$ from thermal fluctuations of the total magnetization $M$ of a certain magnetic system \cite{mc3}.
\begin{figure}[tbp]
\begin{center}
\includegraphics[width=3.5in]{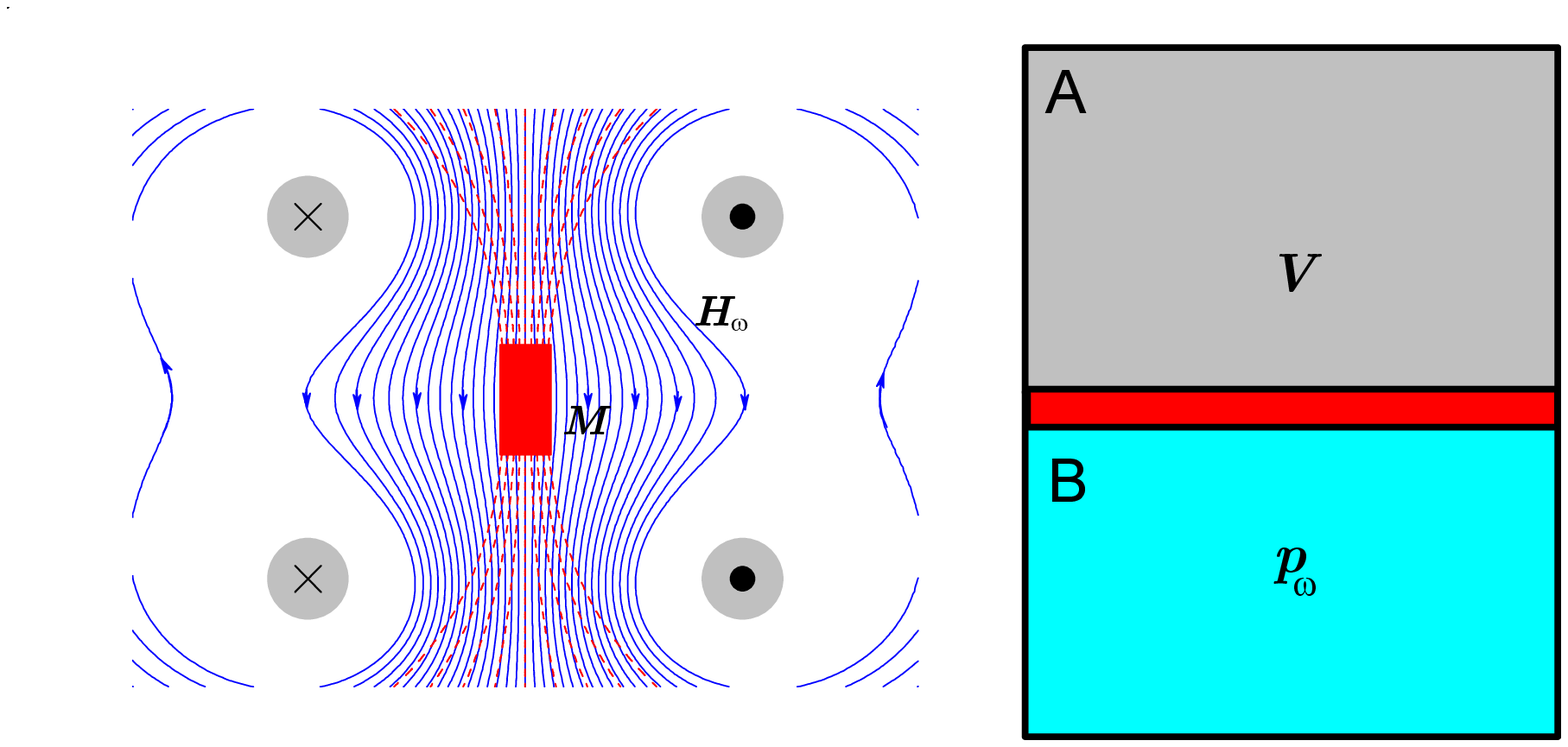}
\end{center}
\caption{(Color online) Correlated thermal fluctuations analogous to the one described by fluctuation relation (\ref{fdr}) can also be observed among other pairs of conjugated thermodynamical variables. Left: Thermal fluctuations of the total magnetization $M$ of a magnetic sample (red rectangle) induce thermal fluctuations in its associated total magnetic flux (red dash lines) through the Helmholtz coils (whose cross sections are represented here by four grey circles). As consequence of electromagnetic induction, the total magnetic field $H_{\omega}$ of these Helmholtz coils (blue lines) experiences correlated thermal fluctuations $\left\langle\delta H_{\omega}\delta M\right\rangle$ with the total magnetization $M$ of the sample. Right: Schematic representation of two finite fluid systems \textbf{A} and \textbf{B} that are separated by an moving wall or piston (red rectangle). The total volume $V$ of fluid system \textbf{A} experiences correlated thermal fluctuations $\left\langle\delta p_{\omega}\delta V\right\rangle$ with the external pressure $p_{\omega}$ of fluid system \textbf{B}.}\label{CorrelationCoil.eps}
\end{figure}
This relation can be generalized as follows:
\begin{eqnarray}\nonumber
\chi_{T}=\beta\left\langle \delta M^{2}\right\rangle-\beta \chi_{T}\left\langle\delta H_{\omega}\delta M\right\rangle+\\
+\left[T\left(\partial M/\partial T\right)_{T}-M\right]\left\langle\delta \beta_{\omega}\delta M\right\rangle,\label{magn.fr}
\end{eqnarray}
while the corresponding fluctuation relation for the heat capacity at constant magnetic field $C_{H}$ is given by:
\begin{eqnarray}\nonumber
C_{H}=\beta^{2}\left\langle \delta Q^{2}\right\rangle+C_{H}\left\langle\delta \beta_{\omega}\delta Q\right\rangle-\\
-\left[T\left(\partial M/\partial T\right)_{T}-M\right]\beta^{2}\left\langle\delta H_{\omega}\delta Q\right\rangle.\label{Ch.fr}
\end{eqnarray}
Here, $\beta_{\omega}$ and $H_{\omega}$ represent the environmental inverse temperature and the intensity of the external magnetic field that is applied over a magnetic system of interest. Moreover, $\delta Q=\delta U-H\delta M$ is the amount of heat absorbed or transferred by the system at the equilibrium, where $\left\langle \delta Q\right\rangle=0$. Under general thermodynamical conditions, all these macroscopic quantities and thermodynamical parameters undergo thermal fluctuations that are coupled among them.

\begin{figure}[tbp]
\begin{center}
\includegraphics[width=3.5in]{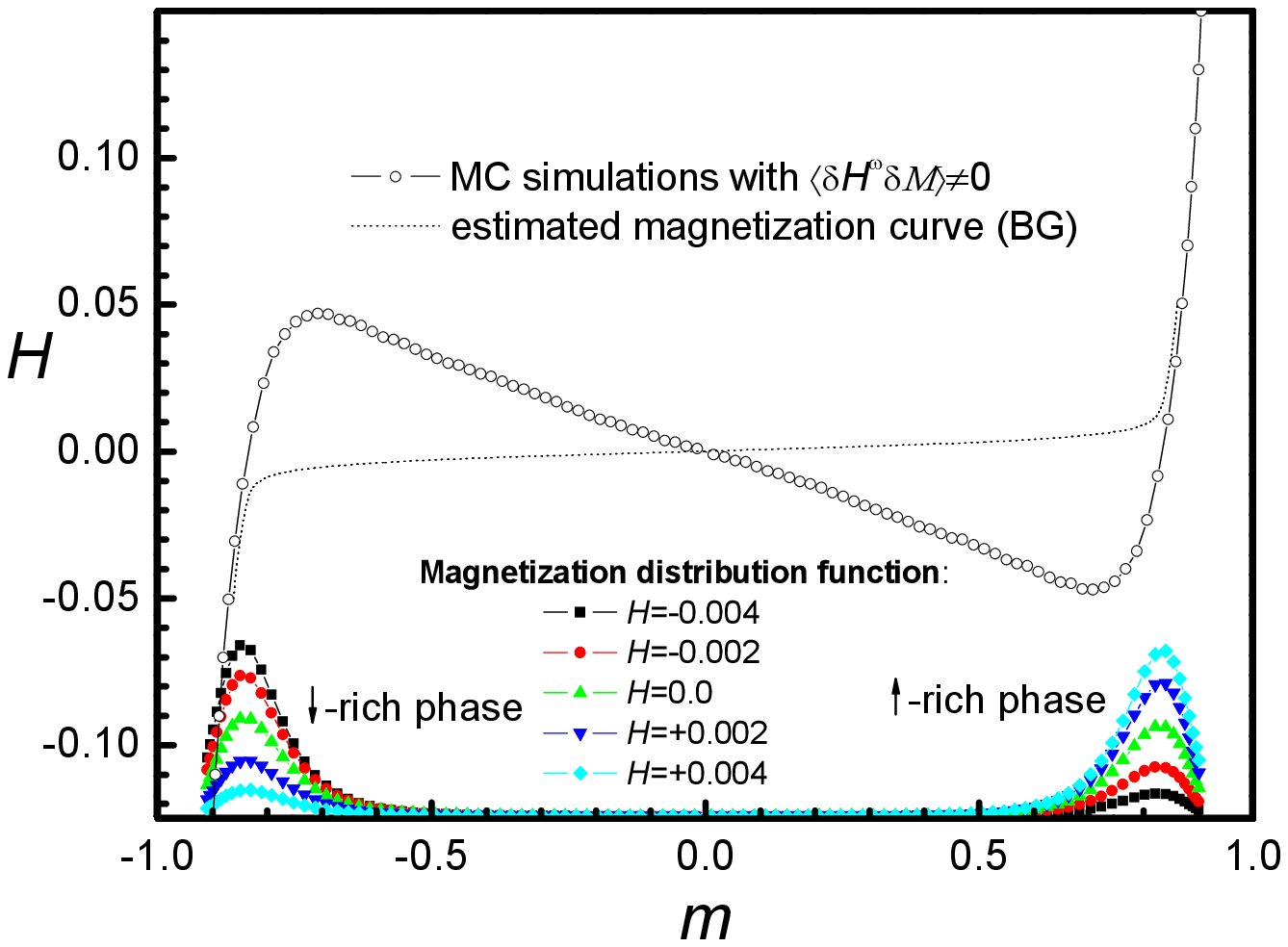}
\end{center}
\caption{(Color online) Behavior of distributions of magnetization per site $m=M/N$ of Ising model on the square lattice with periodic boundary conditions for different constant values of the external magnetic field $H$ at constant temperature $T<T_{c}$, where $T_{c}$ is the temperature critical value of ferro-para continuous PT of this model (after \cite{vel-geft}). Open circles is a MC estimation of microcanonical dependence of magnetization \emph{versus} the external magnetic field (an analogous of microcanonical caloric curve). Doted line represents dependence of average magnetization $\left\langle m\right\rangle$ when the intensity of the external magnetic field $H$ is smoothly varying from negative to positive values at constant temperature. Clearly, this canonical dependence (constant values of control parameters $T$ and $H$) fails to describe the S-bend of its microcanonical counterpart. Moreover, the region with negative values of isothermal magnetic susceptibility $\chi_{T}<0$ is poorly populated by magnetization distributions for constant values of the external magnetic field $H$ and temperature $T$. In fully analogy as the energy-temperature fluctuation relation (\ref{fdr}) enables the study of systems with negative heat capacities $C<0$, fluctuation relation (\ref{magn.fr}) was employed in this MC study to obtain anomalous values of isothermal magnetic susceptibility $\chi_{T}$.}\label{distributions.eps}
\end{figure}

Correlated thermal fluctuations as the ones commented in Fig.\ref{CorrelationCoil.eps} are systematically omitted by conventional ensembles of statistical mechanics, such as canonical ensemble (\ref{can}) and its generalization, the so-called \emph{Boltzmann-Gibbs distributions} \cite{Reichl}. Consequently, its associated fluctuation relations as (\ref{can.fr}) and (\ref{suscept.fr}) are incompatible with the existence of response functions with anomalous values, such as negative heat capacities $C_{H}<0$ or negative isothermal susceptibilities $\chi_{T}<0$ in a ferromagnetic system. Some direct consequences as the inequality (\ref{Thirring.ineq}) also imply a violation of zeroth-law of thermodynamics \cite{vel-emc3,Guggenheim,Ramirez}. In the framework of MC simulations, all these general fluctuation relations are relevant because of the occurrence phase transitions is mostly accompanied with the existence of anomalous values in response functions \cite{gro1}. This connection is also shown in Fig.\ref{distributions.eps} for the case of Ising model on the square-lattice $L\times L$ with periodic boundary conditions, where fluctuation relation (\ref{magn.fr}) was employed to study anomalous values of isothermal magnetic susceptibility $\chi_{T}$ that are found below critical temperature of ferro-para PT of this paradigmatic model system \cite{vel-geft}. All that is discussed in this work concerning to MC study of microcanonical energy-temperature dependence and its associated response function, the microcanonical heat capacity, can directly be extended to other situations with several control parameters introducing appropriate modifications. This perspective was employed in Ref.\cite{vel-geft} to obtain microcanonical magnetization \emph{versus} magnetic field dependence shown in Fig.\ref{distributions.eps}.

\subsection{Extended canonical MC algorithms}

As already commented, the use of a bath with finite heat capacity in MC simulations was firstly proposed by Gerling and H\"{u}ller \cite{Gerling}. These authors considered the system of interest is put in thermal contact with a \emph{bath with constant heat capacity} $C_{\omega}$ (e.g., the system acting as a bath can be an ideal gas). Let us denote by $U_{T}$ the total energy of the system and the bath, which remains fixed when they are put in thermal contact. It can be shown that the inverse temperature of the bath under the above conditions depends on the system energy $U$ as follows:
\begin{equation}
\beta_{\omega}(U)=\frac{C_{\omega}}{U_{T}-U}.
\end{equation}
As naturally expected, this situation is just a particular case among all possible equilibrium situations considered by generalized fluctuation relation (\ref{fdr}). If the system size $N$ is sufficiently large, the thermodynamic influence of every bath with finite heat capacity $C_{\omega}>0$ turns asymptotically equivalent as consequence of applicability of Gaussian approximation for energy fluctuations. However, significant differences in system fluctuating behavior arise when the system size is not so large. In fact, the bath proposed by Gerling and H\"{u}ller is not the most convenient one.

For an arbitrary bath with probability weight $\omega(U)$, its corresponding inverse temperature $\beta_{\omega}(U)$ can be expressed as follows\footnote{This definition follows from combining Einstein postulate $dp(U|U_{T})=A\exp\left[S_{T}(U|U_{T})\right]dU$ of classical fluctuation theory and the additivity of entropy $S_{T}(U|U_{T})=S(U)+S_{\omega}(U_{T}-U)$.} \cite{vel-tur}:
\begin{equation}\label{def.bath}
\beta_{\omega}(U)=-\frac{\partial}{\partial U}\log \omega(U).
\end{equation}
Notice that this definition contains temperature parameter $\beta$ of canonical ensemble as a particular case (\ref{can}). The energy dependence of inverse temperature $\beta_{\omega}(U)$ can be developed in power series around a certain reference energy $U_{s}$ as follows:
\begin{equation}\label{expansion}
\beta_{\omega}(U)=\beta_{s}+\sum^{+\infty}_{n=1} a_{n}\left(U-U_{s}\right)^{n}.
\end{equation}
If thermal fluctuations of the system energy are sufficiently small, in particular, when the size $N$ of the system under study is sufficiently large, high-order terms in power expansion (\ref{expansion}) can be disregarded, except the term corresponding to linear approximation:
\begin{equation}\label{env.temperature}
\beta_{\omega}(U)= \beta_{s}+\lambda_{s}\left(U-U_{s}\right)/N.
\end{equation}
For the sake of convenience, we have identified here $a_{1}\equiv\lambda_{s}/N$. Moreover, additional parameters $U_{s}$ and $\beta_{s}$ can be regarded as roughly estimates of the expectation values $\left\langle U\right\rangle $ and $\left\langle \beta_{\omega}\right\rangle $. Expression (\ref{env.temperature}) is the simplest mathematical dependence for the bath inverse temperature $\beta_{\omega}$ that captures the existence of correlated fluctuations $\left\langle\delta \beta_{\omega}\delta U\right\rangle$ described by fluctuation relation (\ref{fdr}). Hereinafter, we shall assume this dependence is \emph{exact}, that is, let us assume a bath that fulfils this expression.

According to definition (\ref{def.bath}), linear dependence (\ref{env.temperature}) corresponds to the \emph{Gaussian ensemble}:
\begin{equation}
\omega_{G}\left( U|\theta\right) =\exp
\left[ f(\theta)-\phi(U|\theta)\right]
\label{GEns}
\end{equation}
introduced by Challa and Hetherington \cite{Challa,Hetherington,ChallaPRA}, where $\theta\equiv\left(U_{s},\beta_{s},\lambda_{s}\right)$ with parameter $\lambda_{s}\geq 0$, and $\phi(U|\theta)$ is the second-order polynomial function:
\begin{equation}\label{phi}
\phi(U|\theta)=\beta_{s}\left(U-U_{s}\right)+\frac{1}{2N}\lambda_{s}\left(U-U_{s}\right)^{2}.
\end{equation}
Formally speaking, Gaussian ensemble (\ref{GEns}) corresponds to a bath that is composed of an \emph{hypothetical substance} whose heat capacity $C$ depends on its temperature $T$ as $C\propto 1/T^{2}$. This type of dependence is indeed observed in the high-temperature limit of a paramagnetic system \cite{Reichl}. However, Gaussian ensemble (\ref{GEns}) can also be regarded as an \emph{non-physical ensemble} for the purpose of MC simulations. Certainly, there is nothing wrong with this interpretation. Non-physical statistical ensembles are usually considered in MC studies with different purposes, as the case of the so-called \emph{multicanonical ensemble} \cite{mc3}. The use of this generalized statistical ensemble here is fully justified by practical purposes. Gaussian ensemble (\ref{GEns}) contains canonical ensemble (\ref{can}) in the limit $\lambda_{s}\rightarrow0^{+}$, as well as microcanonical ensemble:
\begin{equation}
\omega(U|U_{s})=\frac{1}{\Omega(U_{s})}\delta\left[U-U_{s}\right]
\end{equation}
in the limit $\lambda_{s}\rightarrow+\infty$. This ensemble is easy to combine with any MC algorithm based on canonical ensemble (\ref{can}) regardless its character local or nonlocal \cite{vel-emc3}. The roughly idea is to replace constant temperature parameter $\beta$ of canonical ensemble (\ref{can}) by the \emph{transition inverse temperature}  $\beta^{t}_{ij}=\left[\beta_{\omega}(U_{i})+\beta_{\omega}(U_{j})\right]$ of the initial and final configurations with energies $U_{i}$ and $U_{j}$, respectively. For the case of Metropolis importance sample \cite{metro,Hastings}, its acceptance probability is modified as follows:
\begin{equation}\label{met.aceptance}
W(U_{i}\rightarrow U_{j})=\min\left[1,\exp\left(-\beta^{t}_{ij}\Delta U_{ij}\right)\right],
\end{equation}
where $\Delta U_{ij}=U_{j}-U_{i}$. Implementation of this statistical ensemble for canonical clusters MC algorithms was extensively discussed in subsection II.C of our precedent paper \cite{vel-emc3}. The simple mathematical form of this ensemble makes all analytical developments of the present methodology easier, such as the analysis of detailed balance and the analysis about the incidence of finite size effects \cite{vel-emc3}.

As naturally expected, statistical expectation values of physical quantities are \emph{ensemble-dependent}. To avoid this difficulty, the primary goal of extended canonical MC methods is the calculation \emph{microcanonical quantities} derived from the first-derivatives of the system microcanonical entropy $S(U)$, such as the microcanonical caloric curve $\beta(U)$ (energy dependence of the system inverse temperature) and the curvature curve $\kappa(U)$:
\begin{equation}\label{mic}
\beta(U)=\frac{\partial S(U)}{\partial U}\mbox{ and }\kappa(U)=-N\frac{\partial^{2} S(U)}{\partial U^{2}}.
\end{equation}
This second quantity is directly related to the microcanonical heat capacity $C$ as $\kappa =\beta ^{2}N/C$. In fully analogy with dynamic ensemble MC method \cite{Gerling}, calculation of microcanonical caloric curve can be achieved in the framework of Gaussian approximation of energy distribution function using the expectation values of the bath inverse temperature and the system energy:
\begin{equation}\label{estimation}
\beta_{e}\simeq\left\langle \beta_{\omega}\right\rangle\mbox{ and }U_{e}\simeq\left\langle U\right\rangle,
\end{equation}
where $U_{e}$ represents the most likely value of the system energy. The value of microcanonical curvature $\kappa_{e}=\kappa(U_{e})$ at the energy $U_{e}$ can be estimated from generalized fluctuation relation (\ref{fdr}) as follows:
\begin{equation}\label{kappa.Gauss}
\kappa_{e}\simeq \frac{1-\lambda_{s} \left\langle \delta
U^{2}\right\rangle /N}{\left\langle \delta U^{2}\right\rangle /N}.
\end{equation}
Although the above estimations of microcanonical dependencies (\ref{mic}) are only exact in the thermodynamic limit $N\rightarrow\infty$, the incidence of finite size effects is considerably reduced using the following formulae \cite{vel-emc1}:
\begin{eqnarray}
U_{e} =\left\langle U\right\rangle -\frac{1-\psi _{1}}{2\left\langle \delta
U^{2}\right\rangle }\left\langle \delta U^{3}\right\rangle +O\left( \frac{1}{%
N^{3}}\right) ,  \notag \\
\beta _{e} =\left\langle \beta _{\omega }\right\rangle -\lambda_{s} \frac{1-\psi
_{1}}{2N\left\langle \delta U^{2}\right\rangle }\left\langle \delta
U^{3}\right\rangle +O\left( \frac{1}{N^{3}}\right) ,  \label{EE} \\
\kappa _{e} =\frac{1-\psi _{1}-\lambda_{s} \left\langle \delta
U^{2}\right\rangle /N}{\left\langle \delta U^{2}\right\rangle /N}+O\left(
\frac{1}{N^{2}}\right) . \notag
\end{eqnarray}
Here, $\psi _{1}=\frac{6}{5}\epsilon _{2}+\frac{11}{30}\epsilon _{1}$ is a second-order correction term defined from the cumulants $\epsilon _{1}$ and $\epsilon _{2}$:
\begin{equation}
\epsilon _{1}=\frac{\left\langle \delta U^{3}\right\rangle ^{2}}{%
\left\langle \delta U^{2}\right\rangle ^{3}},~\epsilon _{2}=1-\frac{%
\left\langle \delta U^{4}\right\rangle }{3\left\langle \delta
U^{2}\right\rangle ^{2}}.  \label{EA}
\end{equation}
This same calculations enable us to obtain a roughly estimations for the third and the four-order derivatives of the entropy:
\begin{eqnarray}
\zeta^{3}_{e}&=&N^{2}\frac{\partial^{3}S(U_{e})}{\partial U^{3}}=N^{2}\frac{%
\left\langle \delta U^{3}\right\rangle}{\left\langle \delta
U^{2}\right\rangle ^{3}}\left(1-3\psi_{1}\right) +O\left( \frac{1}{N^{2}}%
\right) ,  \notag \\
\zeta^{4}_{e}&=&N^{3}\frac{\partial^{4}S(U_{e})}{\partial U^{4}}=-\psi_{2}%
\frac{N^{3}}{\left\langle \delta U^{2}\right\rangle ^{3}}+O\left( \frac{1}{N}%
\right),\label{EAA}
\end{eqnarray}
where $\psi_{2}=\frac{12}{5}\epsilon _{2}+\frac{41}{15}\epsilon _{1}$. Ideas behind derivation of this procedure are discussed in Appendix \ref{inference}. Applicability of these formulae is subjected to licitness of \emph{Gaussian approximation} for describing system fluctuating behavior within Gaussian ensemble (\ref{GEns}). This means that its control parameters $(U_{s},\beta_{s},\lambda_{s})$ must be carefully chosen to guarantee applicability of Gaussian approximation.

Roughly speaking, the MC estimation procedure (\ref{estimation}) to obtain microcanonical caloric curve $\beta(U)$ of a given system resembles practical measurements of this dependence. Statistical ensemble that is employed in this type of MC simulation mimics thermodynamical influence of a measuring instrument, e.g., a thermometer. This procedure is always subjected to statistical uncertainties that could be reduced but never eliminated at all \cite{vel-csm}. According to approximation (\ref{kappa.Gauss}), statistical uncertainties for a simultaneous determination energy and its inverse temperature can be estimated in terms of microcanonical curvature $\kappa_{e}$ as follows:
\begin{equation}
\left\langle \delta U^{2}\right\rangle \simeq\frac{N}{\kappa_{e}+\lambda_{s} }%
\mbox{ and }\left\langle \delta\beta_{\omega}^{2}\right\rangle \simeq\frac{1}{N}%
\frac{\lambda^{2}_{s}}{\kappa_{e}+\lambda_{s}}. \label{dispersions}
\end{equation}
Accordingly, statistical uncertainty of the energy can be reduced by increasing the value of parameter $\lambda_{s}$. However, this procedure also implies an increasing of statistical uncertainty of its inverse temperature. Therefore, it is absolutely necessary to establish a compromise between these statistical uncertainties, as example, to minimize the \emph{total dispersion} $\Delta^{2}_{T}$:
\begin{equation}
\Delta^{2}_{T}=\left\langle \frac{1}{N}\delta U^{2}+N\delta\beta_{\omega}^{2}\right\rangle\simeq
\frac{1+\lambda^{2}_{s}}{\kappa_{e}+\lambda_{s}}.
\end{equation}
This criterium leads to the following the optimal value of the control parameter $\lambda_{s}$:
\begin{equation}\label{lambda.opt}
\lambda_{s}=\lambda_{\Delta}\left( \kappa_{e}\right) =\sqrt {1+\kappa_{e}^{2}%
}-\kappa_{e}\mbox{ and }\min(\Delta_{T}^{2})=2\lambda_{\Delta }.
\end{equation}
According to first-order approximation $\delta \beta_{\omega}=\beta^{2}\delta U/C_{\omega}$ employed in derivation of fluctuation relation (\ref{derived.fr}), the parameter $\lambda_{s}$ of Gaussian ensemble (\ref{GEns}) corresponds to the heat capacity $C_{\omega}$ of the bath as $\lambda_{s}\leftrightarrow N\beta^{2}/C_{\omega}$. This way, one obtains the optimal value for the heat capacity $C^{opt}_{\omega}$ of the bath (or the thermometer):
\begin{equation}\label{optimal.heatbath}
C^{opt}_{\omega}=N\beta^{2}\left[\sqrt {1+\left(\frac{N\beta^{2}}{C}\right)^{2}}+\frac{N\beta^{2}}{C}\right]
\end{equation}
that reduces as low as possible the statistical uncertainties during a determination of the microcanonical caloric curve of a given system. It is noteworthy that this last result concerns both its practical determination \cite{vel-efr} as well as its theoretical MC estimation. The fulfilment of this optimization criterium is the best way to force applicability of Gaussian approximation for energy distributions, which is a requirement for the application of point statistical estimation formulae (\ref{EE})-(\ref{EAA}). This criterium also leads to a considerable reduction of finite size effects. This fact is shown in Fig.\ref{gencanonical.eps} for a model system of relative small size. As clearly evidenced, Gaussian-shape of energy distribution is a very good approximation regardless its maximum is located inside the region with negative heat capacities.

\begin{table}[tp]
\centering
\begin{tabular}{ccc}
\hline\hline
\textbf{MC method} & $w _{\tau }$ & $w _{\eta }$ \\ \hline\hline
\multicolumn{1}{c}{\small Metropolis} & $1.06\pm 0.01$ & $1.42\pm 0.01$ \\
\hline
\multicolumn{1}{c}{\small extended Metropolis} & $0.777\pm 0.006$ & $%
0.790\pm 0.008$ \\ \hline
\multicolumn{1}{c}{\small Swendsen-Wang} & $0.432\pm 0.007$ & $0.792\pm
0.008 $ \\ \hline
\multicolumn{1}{c}{\small extended Swendsen-Wang} & $0.098\pm 0.004$ & $%
0.117\pm 0.004$ \\ \hline
\multicolumn{1}{c}{\small Wolff} & $0.474\pm 0.005$ & $0.833\pm 0.007$ \\
\hline
\multicolumn{1}{c}{\small extended Wolff} & $0.094\pm 0.006$ & $0.103\pm
0.006$ \\ \hline\hline
\end{tabular}
\caption{Dynamic critical exponents $w_{\tau}$ and $w_{\eta}$ associated with the size dependencies of
decorrelation time $\tau(N)\propto N^{w_{\tau}}$ and efficiency factor $\eta(N)\propto N^{w_{\eta}}$ at temperature of PT of four-state Potts model on the square lattice $L\times L$ with periodic boundary conditions, with $N=L^{2}$ (after \cite{vel-emc3}).}
\label{critical.exp}
\end{table}

Number $M$ of MC steps that is necessary to reach a convergence of microcanonical caloric curve $\beta(U)$ and the curvature $\kappa(U)$ with an accuracy $N\left\langle\delta \beta^{2}\right\rangle+\left\langle\delta U^{2}\right\rangle/N\leq a^{2}$ and $\left\langle\delta \kappa^{2}\right\rangle<a^{2}$ can be estimated as follows:
\begin{equation}\label{SS}
M\simeq\eta/Na^{2}\mbox{ and }M\simeq2\left(1+\kappa^{2}_{e}\right)\tau/a^{2}.
\end{equation}
where $\tau$ is the decorrelation time and $\eta$ the so-called \emph{efficiency factor} \cite{vel-emc3}:
\begin{equation}\label{efficiency}
\eta=\tau \Delta _{T}^{2}.
\end{equation}
Decorrelation time $\tau$ is the minimum number of MC steps needed to generate effectively independent, identically distributed samples in the Markov chain \cite{mc3}. This quantity crucially depends on the concrete MC algorithm employed in simulations and it is widely regarded as a measure of its efficiency. However, the estimation of microcanonical caloric curve using the present MC methodology is better characterized by the efficiency factor (\ref{efficiency}), which also includes the incidence of the system fluctuating behavior. The simplest way to improve the convergence of a given extended canonical MC algorithm is to minimize the total dispersion $\Delta_{T}^{2}$. As clearly evidenced in Table \ref{critical.exp}, this criterium also involves a sensible improvement of behavior of decorrelation time $\tau$ \cite{vel-emc3}. Since the efficiency factor $\eta$ for a given extended canonical MC method crucially depends on control parameters $\theta=(U_{s},\beta_{s},\lambda_{s})$ of Gaussian ensemble (\ref{GEns}) and the energy value of interest, it is recommendable to employ a variable number $M$ of MC moves for calculating each point estimation of microcanonical dependencies (\ref{mic}).

\subsection{Multicanonical MC methods}

Microcanonical entropy $S(U)$ of a system of interest can be estimated from reweighting MC methods that implements multicanonical ensemble \cite{BergM}, as the case of Wang-Landau method \cite{WangLandau}. Roughly speaking, the essential idea of these MC methods is to carry out a progressive reconstruction of a certain probabilistic weight $\omega_{M}(U)$ that guarantees the existence of a \emph{flat energy histograms}:
\begin{equation}\label{flat}
\omega_{M}(U)W(U)=const,
\end{equation}
which allows a direct estimation $\hat{W}(U)$ of density of states $W(U)$. Once obtained an estimation for microcanonical entropy $\hat{S}(U)=\log \hat{W}(U)$, this information can be employed to calculate any statistical expectation value in any desirable statistical ensemble with probability weight $\omega(U)$ as follows:
\begin{equation}  \label{calc.sev}
\left\langle a\right\rangle=\sum_{U} a(U)\omega(U)\exp\left[\hat{S}(U)\right].
\end{equation}

The many advantages of this type of methodology has been extensively reviewed by Landau and Binder in their book \cite{mc3}: its capacity to enhance \emph{rare events} and obtain a complete information about density states in a single simulation run \cite{WangLandau}; its feasibility to describe systems with complex energy landscapes \cite{Jain,Shell} as well as quantum systems \cite{Troyer1,Troyer2}. A comparison among the present MC methodology and the above reweighting techniques is possible. However, we find more useful to discuss how their different \emph{working principles} could be combined to enhance their respective potentialities. The application of a reweighting technique to improve the accuracy of microcanonical calculations will be discussed in the next section. Therefore, let us restrict here to discuss how arguments employed in the present MC methodology could be employed to improve some aspects of reweighting MC methods.

The point statistical estimates of microcanonical dependencies (\ref{mic}) can be easily employed to provide a \emph{piecewise estimation} for microcanonical entropy $S(U)$ using numerical integration and interpolation methods. This idea was already employed by Viana Lopes and co-workers to develop a progressive piecewise reconstruction of the probabilistic weight of multicanonical ensemble \cite{Lopes}:
\begin{equation}\label{ent.piece}
\omega^{(n)}_{M}(U)=A\exp\left[-S^{(n)}(U)\right].
\end{equation}
Here, $S^{(n)}(U)$ is a \emph{polynomial interpolation} of microcanonical entropy inside a previous explored region $U_{n+1}<U<U_{0}$:
\begin{equation}\label{SS.nint}
S^{(n)}_{VL}(U)=\left\{
\begin{array}{ll}
                   \beta_{0}U, & U>U_{0},\\
                  b_{0}+\beta_{0}\Delta U_{0}-\Delta U_{0}^{2}/2\sigma^{2}_{0} , & U_{1}<U<U_{0},\\
                   \vdots & \vdots \\
                   b_{n}+\beta_{n}\Delta U_{n}-\Delta U_{n}^{2}/2\sigma^{2}_{n}, & U_{n+1}<U<U_{n}, \\
                   b_{n+1}+\beta_{n+1}\Delta U_{n+1}, & U<U_{n+1},
\end{array}
\right.
\end{equation}
plus a \emph{linear extrapolation} outside this region. Here, $\Delta U_{i}\equiv U-U_{i}$, the parameters $(b_{j},\beta_{j},U_{j})$ are obtained as follows:
\begin{eqnarray}\nonumber
&U_{j+1}=U_{j}-\alpha\sigma_{j},\, \beta_{j+1}=\beta_{j}+\alpha/\sigma_{j},&
\\
&b_{j+1}=\beta_{j}U_{j+1}-\alpha^{2}/2,&
\end{eqnarray}
where the step parameter $\alpha\simeq 2-4$. The parameters $\beta_{i}$ and $\sigma_{i}$ are point estimates of microcanonical inverse temperature and the energy statistical dispersion within canonical ensemble at the energy $U_{i}$:
\begin{equation}
\frac{dS(U_{i})}{dU}=\beta_{i}\mbox{ and }-\frac{d^{2}S(U_{i})}{dU^{2}}\simeq\frac{1}{\sigma^{2}_{i}}
\end{equation}
External linear extrapolation in (\ref{SS.nint}) enables the exploration of unknown energy region $U<U_{n+1}$ at constant inverse temperature, which is employed to estimate statistical dispersion $\sigma^{2}_{n+1}$ using the rule:
\begin{equation}\label{rule.sigma}
\sigma^{2}_{n+1}=\left\langle (U-U_{n+1})^{2}\Theta(U_{n+1}-U)\right\rangle,
\end{equation}
with $\Theta(x)$ being Heaviside step function. According to these authors, piecewise estimation (\ref{ent.piece}) and (\ref{SS.nint}) reduces \emph{tunneling times} of multicanonical MC dynamics \cite{Lopes}.

A clear limitation of the above procedure is that the microcanonical entropy $S(U)$ is assumed to be \emph{a concave function everywhere}. This means that this method cannot be applied to systems with negative heat capacities. A simple way to overcome this limitation is to employ the following piecewise formula:
\begin{equation}\label{SS2.nint}
S^{(n)}_{G}(U)=\left\{
\begin{array}{ll}
                   \beta_{0}U, & U>U_{0},\\
                  b_{0}+\beta_{0}\Delta U_{0}-\kappa_{0}\frac{\Delta U_{0}^{2}}{2N} , & U_{1}<U<U_{0},\\
                   \vdots & \vdots \\
                   b_{n}+\beta_{n}\Delta U_{n}-\kappa_{n}\frac{\Delta U_{n}^{2}}{2N}, & U_{n+1}<U<U_{n}, \\
                   b_{n+1}+\phi(U|\theta_{n+1}), & U<U_{n+1},
\end{array}
\right.
\end{equation}
where the use of statistical dispersions $\sigma^{2}_{i}$ was replaced by the microcanonical curvature $\kappa_{i}$. Moreover, linear branch of Eq.(\ref{SS.nint}) for energies $U<U_{n+1}$ is now replaced by the function $\phi(U|\theta_{n+1})$ of Gaussian ensemble (\ref{GEns}) with control parameters $\theta_{n+1}=(U_{n+1},\beta_{n+1},\lambda_{s})$. The optimal value of parameter $\lambda_{s}$ can be estimated from expression (\ref{lambda.opt}) using the previous value of microcanonical curvature $\kappa_{e}\simeq \kappa_{n}$. The values of the energy $U_{n+1}$ and its corresponding mictocanonical inverse temperature $\beta_{n+1}$ can be estimated as follows:
\begin{equation}
U_{n+1}=U_{n}-\alpha\sigma_{n},\, \beta_{n+1}=\beta_{n}+\alpha\kappa_{n}\sigma_{n}/N,
\end{equation}
while the value of constant parameters $b_{j+1}$ are obtained by continuity condition:
\begin{equation}
b_{j+1}=b_{j}+\beta_{j}\Delta_{j}-\kappa_{j}\Delta^{2}_{j}/2N,
\end{equation}
where $\Delta_{j}=-\alpha\sigma_{j}$ and $b_{0}=\beta_{0}U_{0}$. Statistical dispersion $\sigma^{2}_{n+1}$ is also obtained from the rule (\ref{rule.sigma}), which can be employed to estimate microcanonical curvature $\kappa_{n+1}$ using Gaussian approximation:
\begin{equation}
\kappa_{n+1}\simeq N/\sigma^{2}_{n+1}-\lambda_{s}.
\end{equation}
As expected, polynomial interpolation (\ref{SS2.nint}) is now able to describe convex regions of microcanonical entropy. The use of Gaussian ensemble in the unexplored energy region $U<U_{n+1}$ enables the access to regions with negative values of microcanonical curvature curve $\kappa(U)$.

\begin{figure}[tbp]
\begin{center}
\includegraphics[width=3.5in]{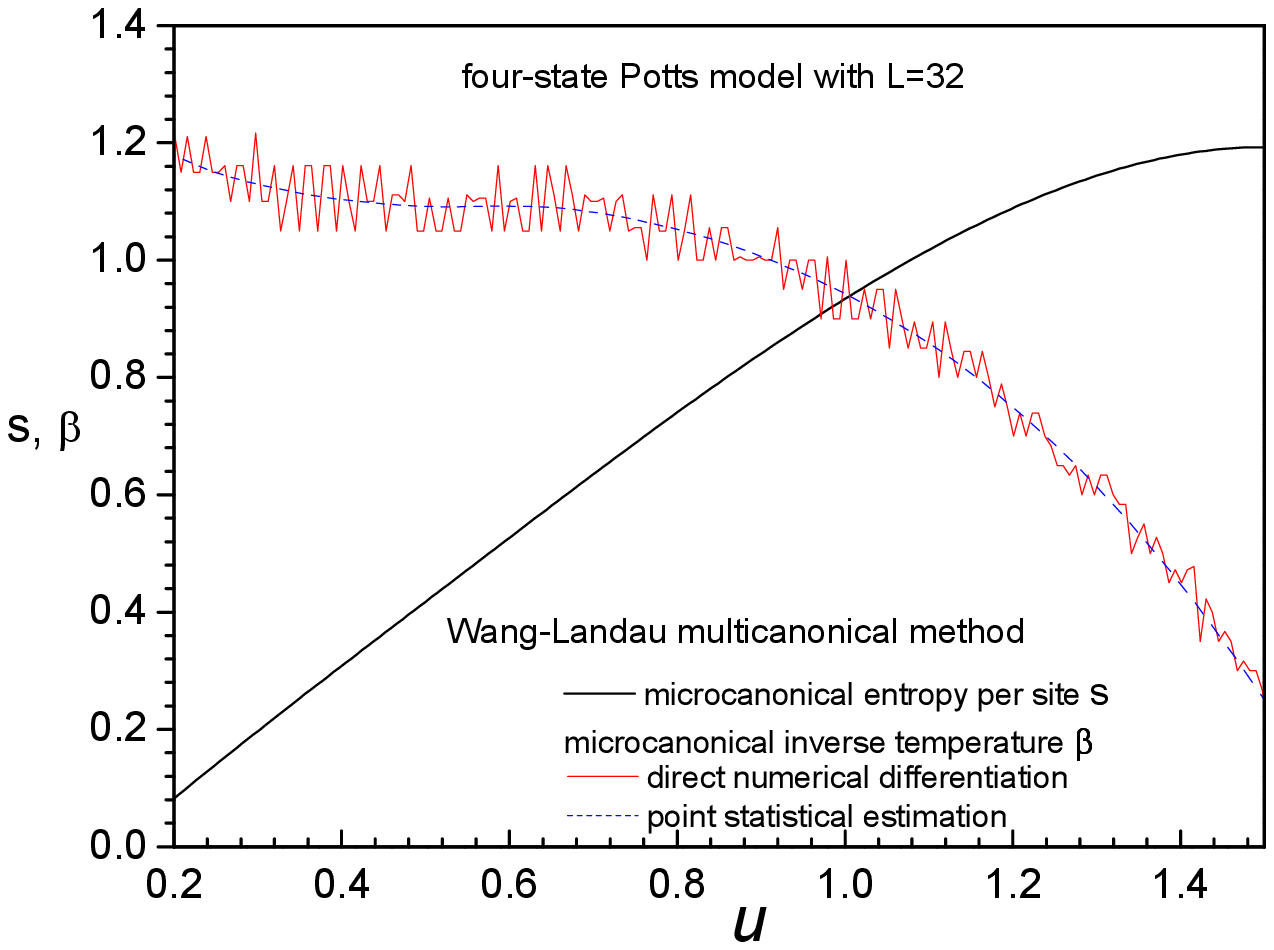}
\end{center}
\caption{(Color online) Entropy per site $s(u)$ and microcanonical inverse temperature $%
\protect\beta(u)$ of 2D four-state Potts model estimated from Wang-Landau
multicanonical algorithm. Here, the variable $u$ denotes the energy per
site, $u=U/N$, with $N=L^{2}$.}\label{Wang-Landau.eps}
\end{figure}

Procedures of numerical integration or interpolation, such as (\ref{calc.sev}) and (\ref{SS2.nint}), do not produce a significant enhancement of statistical uncertainties of any MC estimation of the entropy using reweighting techniques or the point statistical estimation of microcanonical dependencies (\ref{mic}). However, statistical uncertainties turn significant when one is interested on calculation of entropy derivatives using its MC estimation $\hat{S}(U)$. Although they are small, statistical errors introduce considerable affectation during a direct numerical differentiation of entropy estimation $\hat{S}(U)$. A particular demonstration of this problem is shown in Fig.\ref{Wang-Landau.eps}, where entropy estimation $\hat{S}(U)$ of four-state Potts model on the square lattice $32\times32$ obtained from Wang-Landau MC method was employed to estimate microcanonical caloric curve $\hat{\beta}(U)$ by direct numerical differentiation \cite{vel-emc3}.

One can employ different criteria to reduce \emph{roughness} of numerical derivatives, such as adjacent averaging or Savitzky-Golay filter \cite{Savitzky-Golay}. To our knowledge, the previous methods do not follow a specific statistical criteria to deal with a data obtained from MC simulations. We think that a more suitable criteria to obtain smoothly derivatives for MC estimates of entropy $\hat{S}(U)$ is the use of point statistical estimation formulae (\ref{EE})-(\ref{EAA}). Statistical expectation values of this procedure can be obtained from the application of the formula (\ref{calc.sev}). This procedure was already employed in our previous work \cite{vel-emc3}, whose results are also shown here in Fig.\ref{Wang-Landau.eps}. Although these estimations are still affected by incidence of finite size effects, the same ones are very small (see comparative study shown in Fig.3 of Ref.\cite{vel-emc2}). Moreover, the same procedure provides a direct estimation for entropy derivatives of higher-order. Even, one can still obtain better improvements of formulae (\ref{EE})-(\ref{EAA}) by including higher-order correlations of the system fluctuating behavior (see additional comments in Appendix \ref{inference}).

\begin{figure}[tbp]
\begin{center}
\includegraphics[width=3.5in]{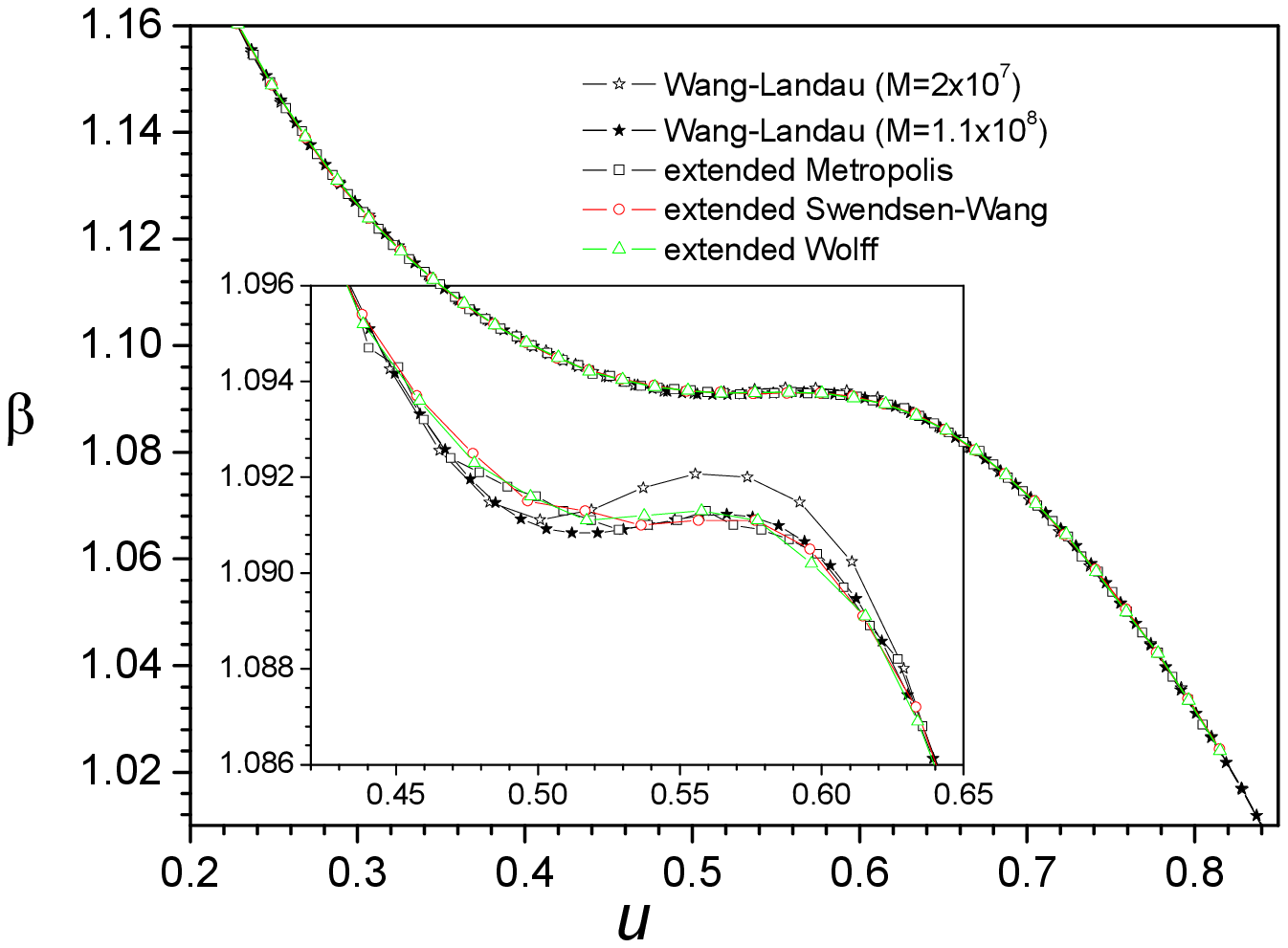}
\end{center}
\caption{(Color online) Microcanonical caloric curve of four-state Potts model system on the square lattice $32\times32$ with periodic boundary conditions, which was estimated from two different realizations of Wang-Landau method and extended versions of canonical MC algorithms of Metropolis importance sample, Wang-Landau and Wolff clusters algorithms (after \cite{vel-emc3}). The agreement among these all these MC methods is very good, although some small discrepancies are clearly evidenced in the inset panel, where these same dependencies were represented with lower energy and inverse temperature scales in order to appreciate better the mathematical behavior of these curves near PT.}\label{compara.previous.eps}
\end{figure}

For comparison purposes, we show in Fig.\ref{compara.previous.eps} different estimations of microcanonical caloric curve of this same model system using the extended versions of canonical MC algorithms of Metropolis importance sampling, and clusters algorithms of Swendsen-Wang and Wolff, as well as two runs of Wang-Landau method of different long\footnote{For implementing Wang-Landau multicanonical method, we have considered a minimum entry of 95\% of the mean value for histogram of energies visited. First simulation run with $M=2\times10^{7}$ steps was extended until parameter $f$ reaches the value $f=\exp(10^{-7})$. Second simulation run with $M=1.1\times10^{8}$ steps was extended until parameter $f$ reaches the value $f=\exp(10^{-8})$.}. According to results shown in the main panel of this figure, the agreement among all these MC method is very good. Nevertheless, one can verify the existence of small discrepancies in the inset panel. In principle, the results obtained from all these MC methods should converge among them. Therefore, the observed discrepancies reveal an insufficient convergence of these MC simulations. It is noteworthy that the existing discrepancies are more significant inside the energy region that contains PT of this model system, which is not a casual fact. According to Eq.(\ref{lambda.opt}) for the minimal total dispersion $\Delta^{2}_{T}$, statistical uncertainties during determination of microcanonical caloric curve $\beta(U)$ are larger where microcanonical curvature curve $\kappa(U)=-N\partial^{2}S(U)/\partial U^{2}$ exhibits its lower values. In other words, statistical uncertainties associated with estimation of microcanonical caloric curve are \emph{nonuniform}.

Extended canonical MC algorithms explore a small energy region in each simulation run because of the use of Gaussian ensemble (\ref{GEns}) with optimal parameters. Consequently, the long of simulations can locally be increased to achieve the necessary accuracy for each energy region. Such a goal can be fulfilled using estimation (\ref{SS}) for the number $M$ of MC steps. The increase the long of simulations using Wang-Landau method involves an increase of number of visits in regions where convergence of point statistical estimations (\ref{EE}) was already achieved. Perhaps, the exigence of \emph{flat energy histograms} (\ref{flat}) should be replaced by other mathematical form that increases the number of visits in those energy regions where microcanonical curvature curve $\kappa(U)$ exhibits its lower values. For example, such a goal can be achieved by the following ansatz:
\begin{equation}
H(U)\propto f_{\kappa}(U)=1+\left[\sqrt{1+\kappa^{2}(U)}-\kappa(U)\right]^{2},
\end{equation}
where $f_{\kappa}(U)$ arises as a \emph{redistribution factor} in the probabilistic weight of multicanonical ensemble:
\begin{equation}
\omega_{\kappa}(U)=A\exp\left[-S(U)\right]f_{\kappa}(U).
\end{equation}
Unfortunately, a complete analysis and implementation of this type of modifications is beyond the scope of the present work. By themselves, these questions deserve a more comprehensive analysis in future works.

\section{Improving accuracy}

\subsection{Application of multi-histograms method}

A main goal of multi-histograms method is the estimation of the number of states $W(U)$. Originally, this method was proposed to extract information of histograms obtained from MC simulations based on canonical ensemble \cite{Ferrenberg}. However, its relevant expressions admit a direct extension for any probability weight. The energy distribution $p_{G}(U|\theta)$ associated with Gaussian ensemble (\ref{GEns}) is given by:
\begin{equation}\label{prob}
p_{G}(U|\theta)=\omega_{G}(U|\theta)W(U).
\end{equation}
Formally, the number of states $W(U)$ is obtained from the energy distribution $p_{G}(U|\theta)$ as follows:
\begin{equation}
W(U)= p_{G}(U|\theta)/\omega_{G}(U|\theta).
\end{equation}
The probability distribution $p_{G}(U|\theta)$ can be estimated using the energy histogram $\hat{p}_{G}(U|\theta)$ of a given simulation:
\begin{equation}
p_{G}(U|\theta)\simeq \hat{p}_{G}(U|\theta)=H(U|\theta)/M,
\end{equation}
where $H(U|\theta)$ is the number of MC moves with final energy $U$, and $M=\sum_{U}H(U|\theta)$ is the total number of MC moves. The energy histogram $\hat{p}_{G}(U|\theta)$ is a random quantity with the following mean and variance \cite{Ferrenberg}:
\begin{equation}\label{random}
\left\langle\hat{p}_{G}(U|\theta)\right\rangle=p_{G}(U|\theta)\mbox{ and }\left\langle\delta\hat{p}^{2}(U|\theta)\right\rangle=p_{G}(U|\theta)/\mathcal{N},
\end{equation}
where $\mathcal{N}=M/\tau$ is the effective number of independent MC moves, with $\tau$ being decorrelation time \cite{mc3}. According to the relative error:
\begin{equation}
\frac{\Delta p_{G}(U|\theta)}{p_{G}(U|\theta)}=\frac{1}{\sqrt{\mathcal{N}p_{G}(U|\theta)}},
\end{equation}
this procedure only allows a reliable estimation of $p_{G}(U|\theta)$ for a small region near most probable energy $U_{e}$. This difficulty is avoided combining the information of independent MC runs with different values of control parameters $\theta$. One can employ the estimator $\hat{W}(U)$:
\begin{equation}
\hat{W}(U)=\mathcal{H}(U)/\mathcal{W}(U)
\end{equation}
for the number of states $W(U)$, while its error can be evaluated as follows:
\begin{equation}
\left\langle\delta W^{2}(U)\right\rangle\simeq\mathcal{H}(U)/\mathcal{W}^{2}(U).
\end{equation}
Here, we have considered the superposition functions of probabilistic weights $\mathcal{W}(U)$:
\begin{equation}\label{weights.function}
\mathcal{W}(U)=\sum_{n}\mathcal{N}_{k}\omega_{G}(U|\theta_{k})
\end{equation}
and the energy histograms $\mathcal{H}(U)$:
\begin{equation}
\mathcal{H}(U)=\sum_{k}\mathcal{N}_{k}\hat{p}_{G}(U|\theta_{k}),
\end{equation}
where $\mathcal{N}_{k}$ is the effective number of independent MC moves for $k$-th simulation run. As expected, normalization function $f_{k}=f(\theta_{k})$ of Gaussian ensemble (\ref{GEns}) with control parameters $\theta_{k}$ should be obtained by self-consistence:
\begin{equation}\label{norma.2}
\exp(-f_{k})=\sum_{U}\exp\left[-\phi(U|\theta_{k})\right]\hat{W}(U).
\end{equation}
Numerical resolution of problem (\ref{norma.2}) can be carried out using some type of \emph{scheme of successive iterations}, such as the one described in Appendix \ref{scheme}.

\begin{figure*}[tbp]
\begin{center}
\includegraphics[width=7.0in]{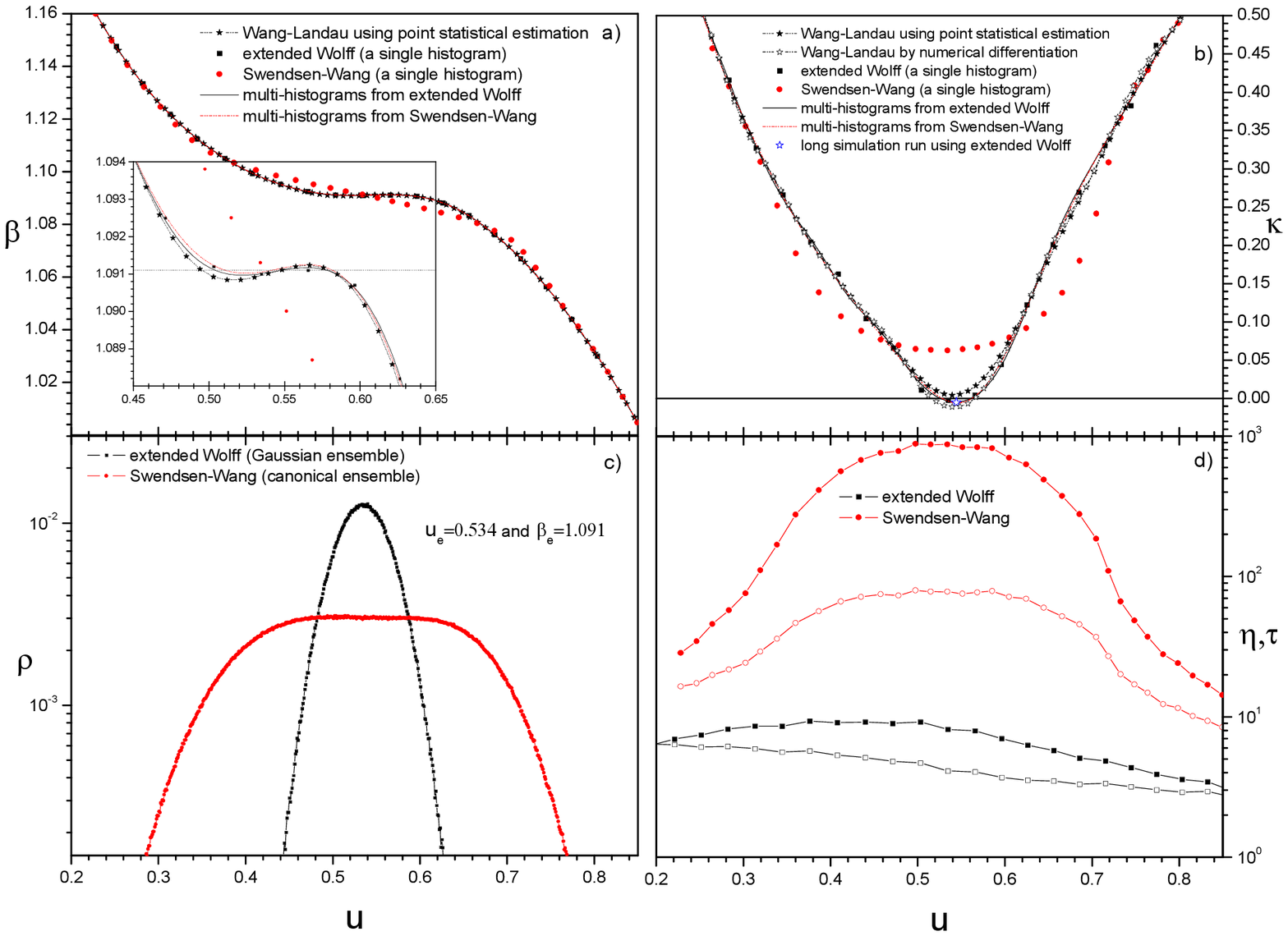}
\end{center}
\caption{(Color online) Comparison among results obtained from MC simulations using extended Wolff and usual Swendsen-Wang clusters algorithms. Results from Wang-Landau method are employed here as reference. Panels a) and b) Microcanonical dependencies $\beta(u)=\partial s(u)/\partial u$ and $\kappa(u)=-\partial^{2} s(u)/\partial u^{2}$ estimated using point statistical estimation (\ref{EE}), where $u=U/N$ and $s=S/N$ are energy and entropy per site, respectively. Squares and circles are the punctual values of these dependencies using a single energy histogram, while solid and dash-dot lines are smooth estimations using re-calculation procedure once obtained microcanonical entropy per site $\hat{s}(u)$ derived from multi-histograms method. To check prediction of multi-histograms method for states with minimum curvature, blue star point corresponds to a point statistical estimation using a single histogram obtained from a very long simulation run with $M=8.1\times10^{7}$ MC steps using extended Wolff algorithm. Additionally, we have included an estimation of microcanonical curvature by direct numerical differentiation of microcanonical caloric curve estimated from Wang-Landau method in panel a). Panel c) Energy histograms near critical point for each MC method. Panel d) Decorrelation time $\tau$ (open squares and circles) and efficiency factor $\eta=\tau \Delta^{2}_{T}$ (solid squares and circles) \emph{versus} the most likely value of energy per particle $u_{e}$ for each simulation run.}\label{comp.WSW.eps}
\end{figure*}

The success of the present methodology relies on a fine tuning of control parameters $(U_{s},\beta_{s},\lambda_{s})$ of Gaussian ensemble (\ref{GEns}). As already commented, their optimal values depend on microcanonical estimates $(U_{e},\beta_{e},\kappa_{e})$, whose calculation is precisely the goal of MC simulation. A practical recipe is to use the microcanonical estimates $(U^{j}_{e},\beta^{j}_{e},\kappa^{j}_{e})$ obtained from a previous MC simulation run, whose energy $U^{i}_{e}$ is close to energy value of interest $U^{j+1}_{e}$. We shall employ the following iterative scheme \cite{vel-emc3}:
\begin{equation}\label{jump}
U^{j+1}_{s}=U^{j}_{e}+\varepsilon_{j};\,\beta^{j+1}_{s}=\beta^{j}_{e}-\kappa^{j}_{e}\varepsilon_{j}\mbox{
and }\lambda^{j+1}_{s}=\lambda_{\Delta}(\kappa^{j}_{e}),
\end{equation}
with $\varepsilon_{j}$ being a variable small energy step. The initial values of the control parameters $(U_{s},\beta_{s},\lambda_{s})$ could be estimated from any canonical MC algorithm far enough from the region of temperature-driven PT. On the other hand, the success of multi-histograms method crucially depends on full coverture of region of interest by energy histograms. To guarantee the overlap between neighboring energy histograms, one can employ the energy dispersion $\Delta U=\sqrt{\left\langle\delta U^{2}\right\rangle}$ of the previous MC simulation, $\varepsilon_{j}=\nu\Delta U_{j}$, where $\nu$ is a fraction in the interval $0<\nu<2$.

\begin{figure*}[tbp]
\begin{center}
\includegraphics[width=7.0in]{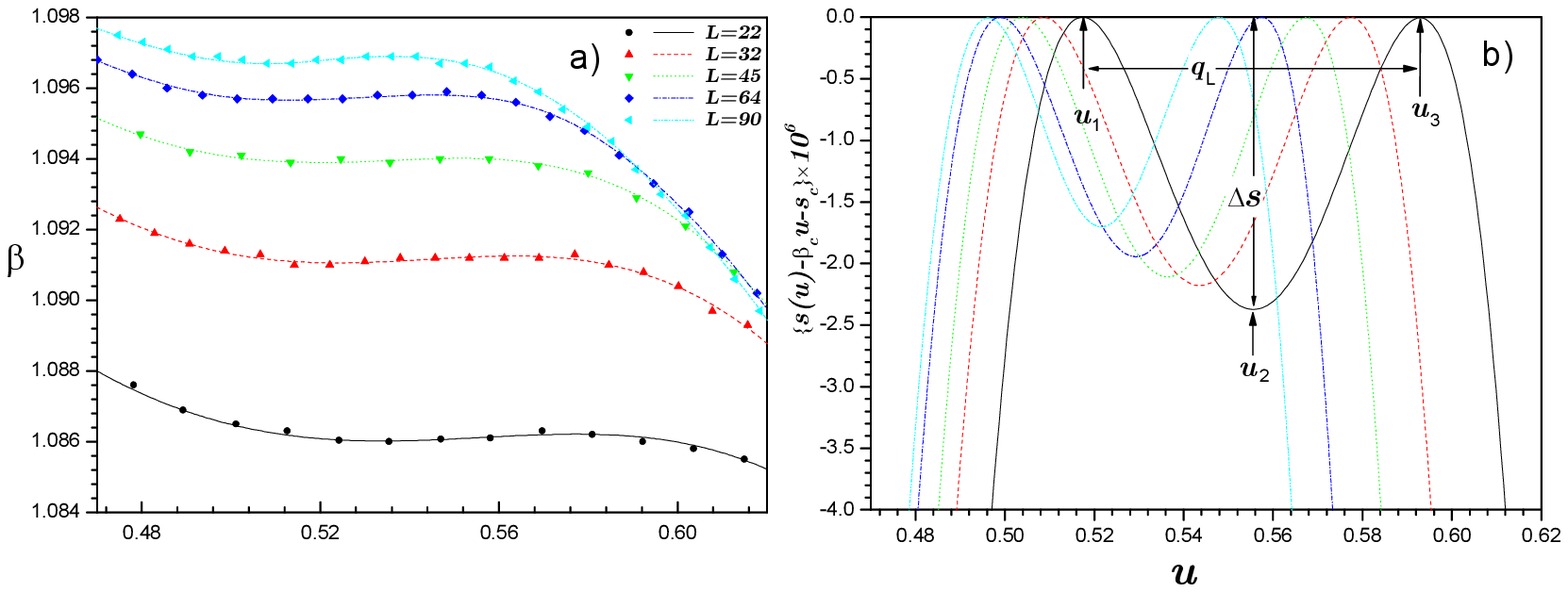}
\end{center}
\caption{(Color online) Some microcanonical dependencies of four-state Potts model on the square lattice $L\times L$ with different $L$, which were estimated considering multi-histograms method combined with extended Wolff algorithm. Panel a) Microcanonical inverse temperature $\beta(u)=\partial s(u)/\partial u$. Panel b) Dependency $s^{*}(u)=s(u)-\beta_{c}u-s_{c}$ obtained from entropy per site $s(u)$ and inverse temperature $\beta_{c}$ corresponding to the temperature driven PT. This auxiliary function reveals better the existence of a \emph{convex intruder} for entropy per site  $s(u)$, which is directly related to the existence of a branch with negative curvature $\kappa(u)=-\partial^{2} s(u)/\partial u^{2}$, or equivalently, states with negative heat capacities $C<0$. Typical energies $(u_{1},u_{2},u_{3})$ and entropy defect $\Delta s=s^{*}(u_{1})-s^{*}(u_{2})$ are employed to characterize this convex intruder region, in particular, to estimate the latent heat per site $q_{L}=u_{3}-u_{1}$.}\label{size.eps}
\end{figure*}

Once obtained the estimation of the number of states $\hat{W}(U)$, microcanonical entropy $S(U)$ can be evaluated using Boltzmann definition $\hat{S}(U)=\log \hat{W}(U)$. The calculation of microcanonical dependencies (\ref{mic}) can be performed using the point statistical estimation formulae (\ref{EE})-(\ref{EA}), where expectation values are evaluated using expression (\ref{calc.sev}). Calculation of microcanonical dependencies (\ref{mic}) demands a good choice of control parameters $\theta$ for each energy. A simple way to achieve this goal is using a simple re-calculation procedure. Essentially, roughly values of microcanonical estimates $(U^{j}_{e},\beta^{j}_{e},\kappa^{j}_{e})$ are considered to provide new values for control parameters $\theta^{j}$:
\begin{equation}\label{recalc.scheme}
U^{j+1}_{s}=U^{j}_{e},\:\beta^{j+1}_{s}=\beta^{j}_{e}\mbox{ and }\lambda^{j+1}_{s}=\lambda_{\Delta}(\kappa^{j}_{e}).
\end{equation}
The control parameters $\theta^{j}$ are employed to provide a new estimation of microcanonical estimates $(U^{j+1}_{e},\beta^{j+1}_{e},\kappa^{j+1}_{e})$. This procedure is repeated until microcanonical estimates reach the convergence with a sufficient accuracy. Final values $(U_{e},\beta_{e},\kappa_{e})$ of this procedure are employed to provide a roughly estimation of control parameters for other energy value of interest using the scheme (\ref{jump}), where energy step $\varepsilon$ is small but arbitrary. This procedure is repeated until obtain a smooth estimation of microcanonical dependencies (\ref{mic}) along energy region of interest.

Maragakis and co-workers have employed in Ref.\cite{Maragakis} a superposition of Gaussian functions similar to expression (\ref{weights.function}) in the framework of Gaussian-mixture umbrella sampling method. However, such a superposition of Gaussians was proposed to estimate a probability distribution using the reweighted statistics from several previous simulations. Result of this fitting procedure is employed to introduce the so-called \emph{biasing potential}. In contrast, superposition of Gaussian functions (\ref{weights.function}) naturally arises here as a consequence of multi-histograms method, specifically, when one combines histograms obtained from independent MC simulations that implement Gaussian ensemble (\ref{GEns}). Its introduction does not involve any fitting procedure because of the number of Gaussian weights and their respective control parameters $\theta_{k}$ were already prefixed by simulations. Only normalization functions $f_{k}$ of Gaussian ensemble (\ref{GEns}) are determined during application of multi-histograms method, but their estimation obeys to a self-consistence requirement (\ref{norma.2}). Gaussian-mixture umbrella sampling was introduced to reconstruct free energy landscapes, while the main purpose of the present methodology is to estimate first derivatives of microcanonical entropy (\ref{mic}).

\subsection{An application example}

We consider the $q$-state Potts model \cite{pottsm}:
\begin{equation}\label{potts}
H=-\sum_{(i,j)}\delta_{\sigma_{i}\sigma_{j}}
\end{equation}
defined on the square lattice $L\times L$ with periodic boundary conditions, where $\sigma_{i}=(1,2,\ldots q)$ is the spin variable of the $i$-th site, while the sum in (\ref{potts}) runs over all nearest neighbors. This family of toy models undergoes both continuous and discontinuous PT at $\beta_{s} = \ln(1 +\sqrt{q})$ in the thermodynamic limit $L\rightarrow\infty$. Their MC study can be performed using different canonical MC algorithms, such as Metropolis importance sample, Swendsen-Wang and Wolff cluster algorithms \cite{SW,pottsm,Wolff}, which enable us to perform a comparative study among them. Additionally, we have also considered Wang-Landau multicanonical MC method \cite{WangLandau}, whose results are employed here as reference to compare with other microcanonical calculations.

To test accuracy of the present improvements of Velazquez and Curilef methodology, let us reconsider the study of the same model system of our previous work: the four-state Potts model \cite{vel-emc3}. According to Baxter exact results \cite{Baxter}, this model undergoes a temperature driven continuous PT at $\beta_{c}\simeq1.0986$ in thermodynamic limit $L\rightarrow+\infty$. For the sake of simplicity, let us restrict this discussion to the cases of extended Wolff clusters algorithm \cite{vel-emc3} and the usual canonical Swendsen-Wang clusters algorithm. We have considered a variable number of MC steps for each calculated point: $M=9.8\times10^{4}\eta$ (extended Wolff) and $M=4.0\times10^{4}\tau$ (Swendsen-Wang), with $\eta$ and $\tau$ being efficiency factor and correlation time a given run, respectively. Typical values for fraction $\nu$ in control parameters scheme (\ref{jump}): $\nu=0.05$ for $L=16-22$ and $\nu=0.5$ for $L=32-90$.

We show in Fig.\ref{comp.WSW.eps} results of MC simulations for the particular case of lattice size $L=32$. We have also included microcanonical estimates obtained from Wang-Landau method using the same data shown in Fig.\ref{compara.previous.eps} for  $M=1.1\times 10^{8}$. According to dependencies shown in panel d) of Fig.\ref{comp.WSW.eps}, extended Wolff algorithm exhibits the lower values of efficiency factor $\eta$ and correlation time $\tau$ for the whole energy region considered in this study. This extended canonical MC algorithm exhibits a greater performance in regard to the usual Swendsen-Wang clusters algorithm. Although canonical ensemble is a particular case of Gaussian ensemble with $\lambda_{s}=0$, any MC methods based on canonical ensemble fails to predict microcanonical dependencies $\beta(u)$ and $\kappa(u)$ near critical point using point statistical estimation (\ref{EE}). This fact is clearly shown in panels a) and b) of this figure. These systematic deviations of microcanonical estimates obtained from Swendsen-Wang MC method relies on the failure of Gaussian approximation of canonical energy distributions near critical point. Such a non-Gaussian behavior of  canonical distributions is observed in energy histograms obtained from Swendsen-Wang MC method, which is shown in panel c) of Fig.\ref{comp.WSW.eps}. On the contrary, Gaussian approximation is fulfilled when one employs Gaussian ensemble (\ref{GEns}) with optimal values of control parameters $\theta=(U_{s},\beta_{s},\lambda_{s})$. This fact is also shown in panel c) of Fig.\ref{comp.WSW.eps} throughout Gaussian-shape of energy distribution obtained from extended Wolff clusters algorithm.

All energy distribution (or histograms) obtained from extended Wolff and usual Swendsen-Wang MC algorithms were combined using multi-histograms method to estimate microcanonical entropy per site $s(u)$. Additionally, we have considered estimation $\hat{s}(u)$ of microcanonical entropy per site obtained from Wang-Landau method. All these estimations were combined with re-calculation procedure to obtain microcanonical dependencies $\beta(u)=\partial s(u)/\partial u$ and $\kappa(u)=-\partial^{2} s(u)/\partial u^{2}$. As clearly evidenced in panels a) and b) of Fig.\ref{comp.WSW.eps}, one observes a fully agreement among microcanonical dependencies obtained from multi-histograms method, the point statistical estimates using extended Wolff algorithm, as well as estimations obtained from Wang-Landau method. According to inset panel of Fig.\ref{comp.WSW.eps}.a, the greater discrepancies among all these MC estimations of microcanonical caloric curve $\beta(u)$ are observed near the inverse temperature of PT, which are of order $\Delta \beta \simeq 10^{-4}$.

Curiously, all these MC estimations are consistent in predicting a \emph{S-bend} of microcanonical caloric curve of this model system outside thermodynamic limit. This mathematical behavior indicates the existence of a small region where microcanonical curvature $\kappa(u)=-\partial^{2} s(u)/\partial u^{2}$ is negative, that is, the existence of an energy region with \emph{negative heat capacities}. Wang-Landau method fails to predict the branch with negative values of microcanonical curvature curve using direct point statistical estimation (\ref{EE}), while its associated microcanonical caloric curve evidences the S-bend. Although the observed deviation is very small, this inconsistence suggests that Wang-Landau estimation of entropy per site $\hat{s}(u)$ does not fulfil the necessary accuracy to obtain a more precise point statistical estimation of microcanonical curvature. In fact, we have obtained a better estimation of this last dependency by applying a direct numerical differentiation on its microcanonical caloric curve $\beta(u)=\partial s(u)/\partial u$. This second procedure now predicts a branch with negative values of microcanonical curvature curve and its results exhibit a better agreement with estimates obtained from multi-histograms method.

As discussed elsewhere \cite{gro1}, the existence of a branch with negative heat capacity is a typical behavior of \emph{finite systems that undergo a temperature driven discontinuous PT}. In fact, this mathematical behavior of microcanonical dependencies is unambiguously observed in all cases of $q$-state Potts models on the square-lattice $L\times L$ with $q>4$ outside thermodynamic limit \cite{vel-emc1,vel-emc2,vel-emc3}. To verify the accuracy of this prediction, we have re-obtained a point statistical estimation of microcanonical inverse temperature and curvature at the energy with minimal value of microcanonical curvature curve. For this purpose, we have considered a single histogram obtained from a very large simulation with $M=8.1\times10^{7}\equiv 8.8\times 10^{6}\eta$ MC steps using extended Wolff algorithm. Control parameters of Gaussian ensemble (\ref{GEns}) for this particular calculation were prefixed using the microcanonical estimates of this notable point, $u_{s}=0.545$, $\beta_{s}=1.0911$ and $\kappa_{s}=-0.0549$, which were previously estimated from multi-histograms method. Point statistical estimation obtained from this new simulation (the blue star point in panel b) of Fig.\ref{comp.WSW.eps}) is in fully agreement with results already obtained from multi-histograms method \footnote{According to results shown in panel d) of Fig.\ref{comp.WSW.eps}, the efficiency factor $\eta$ of extended Wolff algorithm for $L=32$ varies from $2.9$ up to $9.2$ in this energy region. Therefore, the number of steps $M_{k}$ of individual simulations using this clusters algorithm ranges as $M_{k}=(2.8-9.0)\times 10^{5}$ steps, with a total sum $\sum M_{k}=1.4\times10^{7}$. The very large simulation with $M=8.1\times10^{7}$ steps was not considered for calculations using multi-histograms method. This run was only employed to re-calculate microcanonical quantities at the energy with minimum curvature, $u\simeq0.545$.}. According to estimations (\ref{SS}), statistical uncertainties in microcanonical caloric curve are of order $\Delta \beta<10^{-5}$, while the ones of curvature is $\Delta \kappa<3.2\times10^{-4}$. This precision allows us to claim that the existence of this S-bend of microcanonical caloric curve cannot be attributed to a poorly convergence of the data.

\begin{figure}
  \centering
  \includegraphics[width=3.5in]{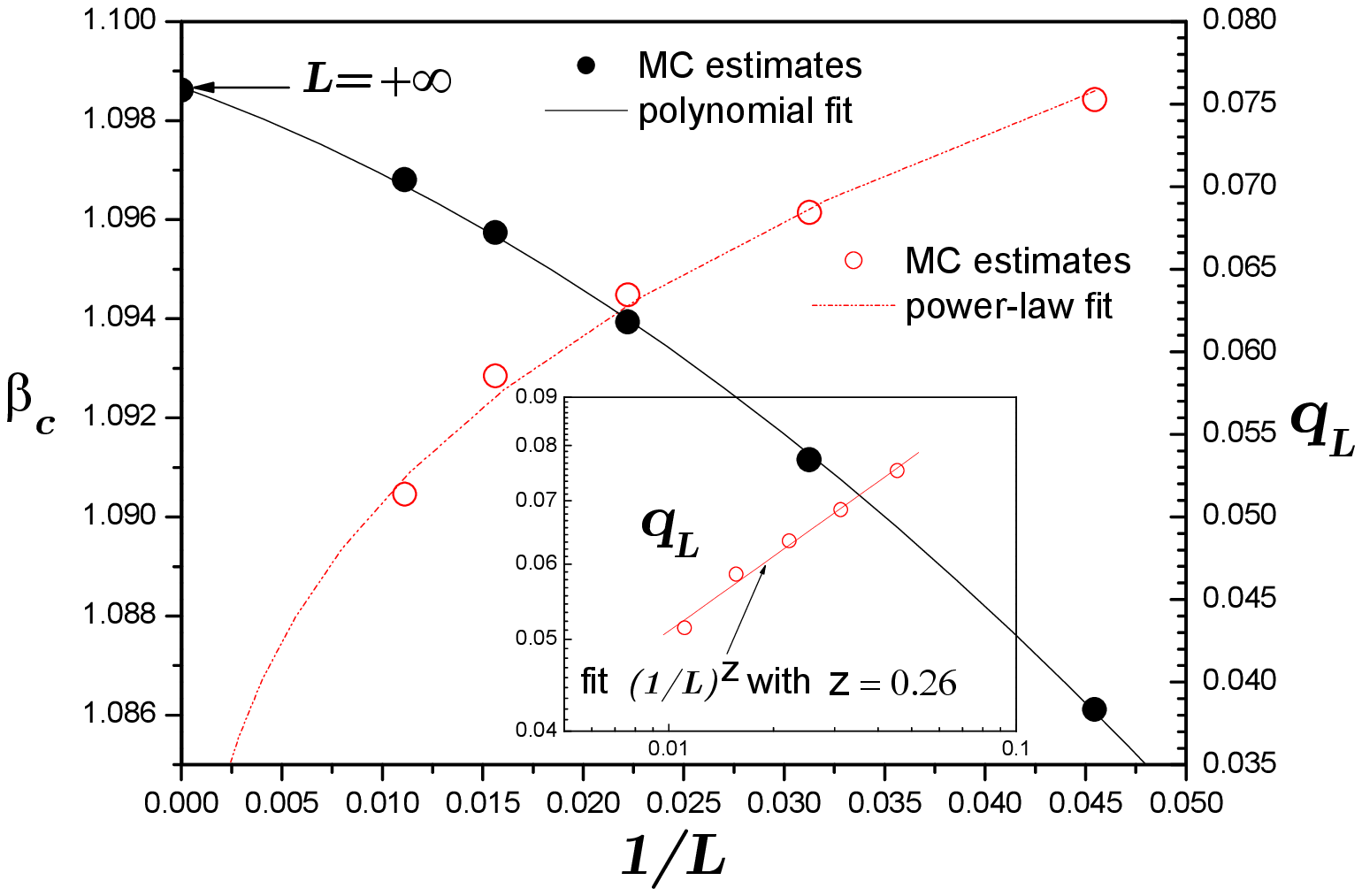}\\
  \caption{(Color online) Dependence of inverse temperature $\beta_{c}$ (black circles) corresponding to temperature driven PT and latent heat per site $q_{L}$ (red open circles) on the inverse lattice size $1/L$. Additionally, we have shown the extrapolation of this data when $1/L\rightarrow0$ using polynomial fits (black line) for the inverse temperature $\beta_{c}(L)$ and a power-law fit for latent heat per site as $q_{L}(L)\propto \left(1/L\right)^{z}$ with $z=0.26\pm0.02$, which is also shown in the inset panel using $\log-\log$ scale.}\label{punctual.eps}
\end{figure}

For a better understanding, microcanonical dependencies (\ref{mic}) were calculated for different values of the lattice size $L$. Because of our modest computational resources, we have restricted here to MC simulations with lattice sizes $L=22-90$  using extended Wolff algorithm and multi-histogram method. Microcanonical dependencies of inverse temperature $\beta(u)$ and entropy per site $s(u)$ are shown in Fig.\ref{size.eps}. Again, these results confirm us the existence of a branch with negative heat capacities in four-state Potts model on the square-lattice $L\times L$ outside thermodynamic limit. As usual, the inverse temperature $\beta_{c}$ corresponding to this type of PT was estimated using Maxwell area rule \cite{Reichl}:
\begin{equation}
\int^{u_{3}}_{u_{1}}\left[\beta(u)-\beta_{c}\right]du=0\rightarrow s(u_{3})-s(u_{1})=\beta_{c}\left(u_{3}-u_{1}\right)
\end{equation}
in conjunction with conditions:
\begin{equation}
\beta(u_{1})=\beta(u_{2})=\beta(u_{3})=\beta_{c}.
\end{equation}
Actually, dependence of entropy per site $s(u)$ was replaced in Fig.\ref{punctual.eps} by the auxiliary function $s^{*}(u)=s(u)-\beta_{c}u-s_{c}$, where $s_{c}$ is suitable constant. This auxiliary function reveals the existence of a \emph{convex intruder} of microcanonical entropy per site $s(u)$. This energy region of convexity can be characterized by the three relevant energies $(u_{1},u_{2},u_{3})$ and \emph{entropy defect} $\Delta s=s^{*}(u_{1})-s^{*}(u_{2})$. The latent heat per site $q_{L}$ is evaluated as $q_{L}=u_{3}-u_{1}$. These notable values are reported in Table \ref{SizeTable}. Size dependencies of inverse temperature $\beta_{c}$ of the PT and latent heat per site $q_{L}$ are shown in Fig.\ref{punctual.eps}.

\begin{table}[tbp] \centering
\begin{tabular}{ccccccc}
\hline\hline
$L$ & $\beta _{c}$ & $u_{1}$ & $u_{2}$ & $u_{3}$ & $q_{L}$ & $\Delta s\times 10^{6}$ \\
\hline\hline
\multicolumn{1}{l}{$22$} & \multicolumn{1}{l}{$1$.$0861$} &
\multicolumn{1}{l}{$0$.$5176$} & \multicolumn{1}{l}{$0$.$5553$} &
\multicolumn{1}{l}{$0$.$5929$} & \multicolumn{1}{l}{$0$.$075$} &
\multicolumn{1}{c}{$2$.$37$} \\
\multicolumn{1}{l}{$32$} & \multicolumn{1}{l}{$1$.$0912$} &
\multicolumn{1}{l}{$0$.$5088$} & \multicolumn{1}{l}{$0$.$5437$} &
\multicolumn{1}{l}{$0$.$5772$} & \multicolumn{1}{l}{$0$.$068$} &
\multicolumn{1}{c}{$2$.$18$} \\
\multicolumn{1}{l}{$45$} & \multicolumn{1}{l}{$1$.$0939$} &
\multicolumn{1}{l}{$0$.$5040$} & \multicolumn{1}{l}{$0$.$5365$} &
\multicolumn{1}{l}{$0$.$5675$} & \multicolumn{1}{l}{$0$.$063$} &
\multicolumn{1}{c}{$2$.$11$} \\
\multicolumn{1}{l}{$64$} & \multicolumn{1}{l}{$1$.$0957$} &
\multicolumn{1}{l}{$0$.$4990$} & \multicolumn{1}{l}{$0$.$5293$} &
\multicolumn{1}{l}{$0$.$5575$} & \multicolumn{1}{l}{$0$.$058$} &
\multicolumn{1}{c}{$1$.$94$} \\
\multicolumn{1}{l}{$90$} & \multicolumn{1}{l}{$1$.$0968$} &
\multicolumn{1}{l}{$0$.$4963$} & \multicolumn{1}{l}{$0$.$5216$} &
\multicolumn{1}{l}{$0$.$5477$} & \multicolumn{1}{l}{$0$.$051$} &
\multicolumn{1}{c}{$1$.$70$} \\
\multicolumn{1}{l}{$\infty $} & \multicolumn{1}{l}{$1$.$0986$} &
\multicolumn{1}{c}{$-$} & \multicolumn{1}{c}{$-$} & \multicolumn{1}{c}{$-$} &
\multicolumn{1}{l}{$0$.$000$} & \multicolumn{1}{c}{$0$.$00$} \\ \hline\hline
\end{tabular}%
\caption{Size dependence of some thermodynamic parameters that characterize the \emph{convex intruder} of entropy per site $s(u)$ shown in panel b) of Fig.\ref{size.eps}.}\label{SizeTable}
\end{table}

At first glance, the present results are quite confusing. Baxter have demonstrated in the past \cite{Baxter} that latent heat of this model system vanishes. However, one can realize that there is no contradiction. Baxter exact result only concerns to four-state Potts model on the square-lattice in \emph{thermodynamic limit} $L\rightarrow+\infty$. By itself, this result does not forbid the existence of macrostates with \emph{negative heat capacities} for finite systems as the cases analyzed in this MC study. In fact, monotonous decreasing of the latent heat per site $q_{L}$ is compatible with an eventual vanishing of this quantity when $L\rightarrow+\infty$. Our MC estimations of latent heat per site $q_{L}$ are consistent with a power-law dependence $q_{L}(L)\propto \left(1/L\right)^{z}$ with $z=0.26\pm0.02$. Of course, it would be desirable to extend the present microcanonical MC estimations for systems with larger lattice sizes $L>90$, which is beyond our computational capability.

As already demonstrated by Baxter himself \cite{Baxter}, four-state Potts model on the infinite square lattice is a \emph{marginal case}: cases with $q>4$ exhibits a temperature driven discontinuous PT, while cases with $q\leq 4$ undergo a continuous PT. According to our results, ambiguities in some behaviors can appear for the marginal case $q=4$ outside thermodynamic limit. For example, multimodal character of canonical energy distributions during phase coexistence phenomenon (see example in Fig.\ref{potts.eps}) leads to an exponential dependence of decorrelation times $\tau(N)\propto \exp( \gamma N)$ with system size $N=L^{2}$ during MC simulations \cite{mc3}. For the particular case of four-state Potts model on the square-lattice $L\times L$, canonical MC algorithms exhibit a power-law dependency of decorrelation times $\tau(N)\propto N^{w_{\tau}}$, whose critical exponents $w_{\tau}$ were already shown in Table \ref{critical.exp}. As expected, such a power-law dependency of decorrelation times is a typical behavior of finite systems at critical temperature of continuous PT \cite{mc3}. According to our results, non-Gaussian form of canonical energy distribution at transition inverse temperature $\beta_{c}$, as the one shown in panel c) of Fig.\ref{comp.WSW.eps}, is explained by the superposition of two close Gaussian peaks. The widths of these peaks are sufficiently large to hide the existence of a bimodal character of energy histogram within canonical ensemble. This behavior cannot be distinguished in canonical energy distribution of this figure because of defect $\Delta s$ of entropy convex intruder is very small. The proximity of these peaks is the reason why canonical MC algorithms do not follow an exponential dependence of decorrelation time $\tau(N)\propto\exp( \gamma N)$.

Barkema and de Boer presented in the past \cite{Barkema} an interesting Monte Carlo study about a dynamical model with parameters $(d^{*},q^{*})$ that resembles $d$-dimensional $q$-state Potts models for non-integer values. Curiously, these authors also reported a non-vanishing latent heat per site $q_{L}$ for the case $d^{*}=2$ and $q^{*}=4$ considering MC simulations with lattice size $L=128$. Their estimated value $q_{L}=0.019$ seems to be compatible with the present study \footnote{A simple extrapolation of numerical results of Table \ref{SizeTable} using power law $q_{L}(L)\propto \left(1/L\right)^{z}$ suggests the value $q_{L}\simeq0.048$ for $L=128$.}. However, these authors do not enter to analyze this particular finding because of they were more interested on behavior of latent heat for non-integer values of parameter $q^{*}$ in thermodynamic limit.

\section{Final remarks}

We have combined the extended canonical MC algorithms with multi-histograms method \cite{Ferrenberg}, which enable us to improve accuracy of microcanonical calculations using point statistical estimation formulae (\ref{EE})-(\ref{EAA}). The resulting technique is sufficiently accurate to detect subtle thermodynamical behaviors during MC simulations. As example of application, we have applied this method to reveals the existence of a very small latent heat during occurrence of temperature driven PT of four-state Potts model on the square lattice $L\times L$ outside thermodynamic limit. Our MC estimates of latent heat per site $q_{L}$ are consistent with a power-law dependence $q_{L}(L)\propto \left(1/L\right)^{z}$ with $z=0.26\pm0.02$, which predicts a vanishing of this quantity when $L\rightarrow+\infty$. Accordingly, the present results are compatible with Baxter exact result about continuous character of temperature-driven phase transition of this model in the thermodynamic limit $L\rightarrow+\infty$.

Velazquez and Curilef methodology \cite{vel-emc1,vel-emc2,vel-emc3} admits other improvements to increase the performance of extended canonical MC methods. A next step is the combination with rejection-free algorithms \cite{Liu}. If possible, resulting algorithms could exhibit much greater performance. This methodology can also be extended to perform a MC study of systems with several control parameters besides energy and temperature. An important step to achieve this purpose was already done in Ref.\cite{vel-geft}, where equilibrium fluctuation relation (\ref{fdr}) was extended to situations with several thermodynamic variables. As already discussed in this work, some arguments of this methodology could be useful to enhance potentialities of other MC methods, such as multicanonical method and its variants \cite{BergM,WangLandau,WangJStat}. Some of these questions will be discussed in forthcoming works.

\begin{acknowledgments}
Velazquez thanks partial financial support of this research from FONDECYT \textbf{1130984} and CONICYT-\textbf{ACT1204} (Chilean agencies). Authors thank to professor A. Zarate because of the access to computational facilities of Research Group on Science Materials and Nanotechnology-UCN.
\end{acknowledgments}

\appendix
\section{Additional discussions}
\subsection{About point statistical estimation}\label{inference}
Formally speaking, point statistical estimation (\ref{EE})-(\ref{EAA}) is an \emph{inference procedure} to determine best guess for first entropy derivatives \cite{Lehmann}. To fix some ideas, let us consider an energy histogram $H(U)$ obtained from a MC simulation based on the Gaussian ensemble (\ref{GEns}):
\begin{equation}
H(U)\propto \omega_{G}(U|\theta)\exp\left[S(U)\right].
\end{equation}
Entropy difference $S(U)-S(U_{e})$ around the most likely value of energy $U_{e}$ can be approximated by the following polynomial:
\begin{eqnarray}
P_{e}(U)=\beta_{e}\Delta U_{e}-\kappa_{e}\frac{\Delta U_{e}^{2}}{2N}+\zeta^{3}_{e}\frac{\Delta U_{e}^{3}}{6N^{2}}
+\zeta^{4}_{e}\frac{\Delta U_{e}^{4}}{24N^{3}}\label{poly.entrop}
\end{eqnarray}
with $\Delta U_{e}\equiv U-U_{e}$, which is Taylor power-expansion of entropy difference up to four-order of approximation. By definition, the energy $U_{e}$ obeys the \emph{stationary condition}:
\begin{equation}\label{constraint.beta}
\beta_{\omega}(U_{e})=\beta_{e},
\end{equation}
where $\beta_{\omega}(U)$ is given by the linear ansatz of Gaussian ensemble (\ref{env.temperature}). Accordingly, the microcanonical inverse temperature parameter $\beta_{e}$ is fully determined by the knowledge of the energy $U_{e}$. Mathematical form of energy histograms $H(U)$ can be approximated by the following distribution:
\begin{equation}\label{PDF}
H(U)\simeq A(\theta,\chi_{e})\exp\left[-Q(U|\theta,\chi_{e})\right],
\end{equation}
where $A(\theta,\chi_{e})$ is a normalization constant and $Q(U|\theta,\chi_{e})$ is the four-order polynomial:
\begin{equation}
Q(U|\theta,\chi_{e})=\phi(U|\theta)-P_{e}(U).
\end{equation}
As naturally expected, parametric distribution (\ref{PDF}) improves Gaussian approximation of energy distributions by including finite size $1/N$-effects. The unknown microcanonical parameters $\chi_{e}=(U_{e},\kappa_{e},\zeta^{3}_{e},\zeta^{4}_{e})$ can be obtained using suitable \emph{estimators} \cite{Lehmann}. In particular, point statistical estimation formulae (\ref{EE})-(\ref{EAA}) follows from the application of the known \emph{method of moments} combined with a perturbative $1/N$-expansion. The idea is to perform calculation of energy moments of $n$-order:
\begin{equation}
\mu_{n}=\mathbf{E}\left(U^{n}\right)=\frac{\sum_{U} U^{n}\exp\left[-Q(U|\theta,\chi_{e})\right]}{\sum_{U} \exp\left[-Q(U|\theta,\chi_{e})\right]}=f_{n}(\chi_{e}|\theta)
\end{equation}
with $n=1-4$. Afterwards, the concrete analytical expressions of functions $f_{n}(\chi_{e}|\theta)$ are inverted as follows:
\begin{equation}
\mu_{n}=f_{n}(\chi_{e}|\theta) \rightarrow \chi_{e}=g_{e}(\mu_{1},\mu_{2},\mu_{3},\mu_{4}|\theta).
\end{equation}
Finally, the estimators $\hat{\chi}_{e}$ of microcanonical parameters $\chi_{e}$ are obtained replacing $\mu_{n}$ by the sample moments:
\begin{equation}
\hat{\mu}_{n}=\frac{\sum_{U} U^{n}H(U)}{\sum_{U} H(U)}.
\end{equation}
Further details about this procedure are discussed in the appendix of Ref.\cite{vel-emc2}.

\subsection{Iterative scheme}\label{scheme}

Firstly, it is convenient to notice that normalization functions $f_{k}$'s in self-consistent problem (\ref{norma.2}) are undetermined by an additive term. If the set of values $f=\left\{f_{k}\right\}$ represents a solution of this problem, the set $f^{*}=\left\{f^{*}_{k}\right\}$ with $f^{*}_{k}=f_{k}+C$ also represent a solution. This fact implies that the estimator $\hat{W}(U)$ is undetermined by a constant factor, $\hat{W}^{*}(U)=\hat{W}(U)\exp(-C)$. This arbitrariness is not a problem because of only entropy change $S(U)=\log W(U)$ for different energies is thermodynamically relevant. Anyway, we shall impose the following constraint:
\begin{equation}\label{sum.const}
\sum_{k}f_{k}=0
\end{equation}
to avoid this arbitrariness. Self-consistent problem (\ref{norma.2}) is solved in this work using the following \emph{scheme of successive iterations}:
\begin{enumerate}
  \item A roughly estimation $f^{n}=\left\{f^{n}_{k}\right\}$ is employed to obtain an estimation for number of states $\hat{W}^{n}(U)$.
  \item A tentative set of values $\check{f}^{n}=\left\{\check{f}^{n}_{k}\right\}$ is obtained from mormalization condition:
\begin{equation}\label{norma.3}
\exp(-\check{f}^{n}_{k})=\sum_{U}A_{k}\exp\left[-\phi(U|\theta_{k})\right]\hat{W}^{n}(U).
\end{equation}
  \item The set $\check{f}^{n}$ is displaced as follows:
\begin{equation}
\tilde{f}^{n}_{k}=\check{f}^{n}_{k}-\frac{1}{Q}\sum_{n}\check{f}^{n}_{n}
\end{equation}
to guarantee imposition of constraint (\ref{sum.const}), with $Q$ being the number of histograms.
  \item A new approximation $f^{n+1}=\left\{f^{n+1}_{k}\right\}$ is obtained as follows:
\begin{equation}
f^{n+1}_{k}=f^{n}_{k}+\epsilon (\tilde{f}^{n}_{k}-f^{n}_{k}),
\end{equation}
where $\epsilon$ is a small positive number.
\end{enumerate}
The present iterative scheme is repeated until the convergence error $\delta_{n}$:
\begin{equation}
\delta_{n}=\sqrt{\frac{1}{Q}\sum_{k}(\tilde{f}^{n}_{k}-f^{n}_{k})^{2}}
\end{equation}
reaches a desirable accuracy. Typically, we have employed the values $\epsilon=0.1$ and $\delta_{n}<10^{-6}$.

\end{document}